\documentclass[a4paper,12pt]{article}

\usepackage{bbm}
\usepackage[utf8]{inputenc}
\usepackage{placeins}
\usepackage{amssymb,amsmath,amsfonts}
\usepackage[colorlinks,linkcolor=purple,citecolor=teal]{hyperref}
\usepackage{dsfont}
\usepackage{color}
\usepackage{graphicx}
\usepackage{hyphenat}
\usepackage{wrapfig}
\usepackage{empheq}
\usepackage{textcomp}
\usepackage[caption=false]{subfig}
\usepackage{rotating}
\usepackage{wrapfig}
\usepackage{pdfpages}
\usepackage{verbatim}
\usepackage{setspace}
\usepackage{array}
\usepackage{caption}
\usepackage[pdf]{pstricks}
\usepackage{upgreek}
\usepackage{cmll}
\usepackage{latexsym}
\usepackage{braket}
\usepackage{cite}
\usepackage{epsfig}
\usepackage[left=2.4cm,top=3.3cm,right=2.4cm,bottom=3.3cm,bindingoffset=0cm]{geometry}
\usepackage[titletoc,page]{appendix}
\usepackage{multicol}
\usepackage{youngtab}

\newcommand{\beq}{\begin{eqnarray}}
\newcommand{\eeq}{\end{eqnarray}}
\newcommand{\bea}{\begin{eqnarray}}
\newcommand{\eea}{\end{eqnarray}}
\newcommand{\be}{\begin{equation}}
\newcommand{\ee}{\end{equation}}

\def\scalar#1#2{\langle{#1}|{#2}\rangle}
\def\brc{\langle}
\def\ckt{\rangle}
\def\const{{\rm const}}
\def\de{\partial}

\numberwithin{equation}{section}

\numberwithin{equation}{section}

\begin{document}

\title{
 \vskip 20pt
{\huge   
Newton's equations  from
quantum mechanics \\
for a macroscopic body in the vacuum
}}


\author{ 
 Kenichi Konishi $^{(1,2)}$  
  \\[13pt]
{\em \footnotesize
$^{(1)}$Department of Physics ``E. Fermi", University of Pisa,}\\[-5pt]
{\em \footnotesize
Largo Pontecorvo, 3, Ed. C, 56127 Pisa, Italy}\\[2pt]
{\em \footnotesize
$^{(2)}$INFN, Sezione di Pisa,    
Largo Pontecorvo, 3, Ed. C, 56127 Pisa, Italy}\\[2pt]
\\[1pt] 
{ \footnotesize  kenichi.konishi@unipi.it  }  
}
\date{}

\vskip 6pt

\maketitle

\begin{abstract}

Newton's force law  $\frac{d {\bf P}}{dt} = {\bf F}$    is derived from the Schr\"odinger equation for   isolated macroscopic bodies,  composite states of  e.g., $N\sim 10^{25},  10^{51}, \ldots$ atoms and molecules,   at finite body temperatures.   We first review three aspects  of quantum mechanics (QM)  in this context:  
  (i)   Heisenberg's uncertainty relations for their center of mass (CM), (ii)  the diffusion of the CM  wave packet,  and (iii) a finite body-temperature which implies a  metastable (mixed-) state of the body:  photon emissions and self-decoherence.  They explain the origin of  the classical trajectory for a macroscopic body.
     The ratio between the range $R_q$ over which the quantum fluctuations of its CM are effective,  and the body's (linear) size  $L_0$,  
   $R_q /L_0  \lesssim  1$ or $R_q/ L_0  \gg 1$,  
tells whether the body's CM behaves classically or quantum mechanically,  respectively.  In the first case,
  Newton's force law  for its CM  follows  from the Ehrenfest theorem. We illustrate this for weak gravitational forces,  a harmonic-oscillator  potential,
    and for  constant external electromagnetic fields slowly varying in space. The derivation of  the canonical Hamilton equations for many-body systems is  also discussed.  
   Effects due to the body's finite size such as the gravitational tidal forces appear 
    in perturbation theory.
Our work  is consistent with the well-known idea  that the emergence of classical physics in  QM   is due  to  the environment-induced decoherence,    but  complements  and completes it, by clarifying the conditions under which Newton's equations follow from QM, and by deriving them explicitly.

\end{abstract}


\newpage

\tableofcontents


\bigskip

\newpage

\section{Introduction   \label{Intro}}

A well-known idea on how the classical behavior of macroscopic bodies  emerges from QM  (QM)   is the so-called environment-induced 
decoherence  \cite{Joos1}-\cite{Arndt1}.  
Due to the scatterings with the environment particles (any kind of external fluxes, light, air molecules, cosmic ray background, etc.) and by the ensuing uncontrolled entanglement with the scattered particles, the macroscopic bodies under study  become mixed states, and lose the capacity of  forming 
coherent quantum superposition of states.  For instance,  a  ``state" describing a split wave packet with two narrow peaks centered at ${\bf r}_A$ and   ${\bf r}_B$, 
\be   |\Psi \ckt   = c_A  |  \Psi_A \ckt  +   c_B   | \Psi_B \ckt\;, \label{Decohrence1}
\ee
is described by the density matrix
 \bea     \rho({\bf r}, {\bf r}^{\prime}) &=&   \Psi({\bf r}) \Psi({\bf r}^{\prime})^*   \nonumber     \\
&=&    |c_A|^2   \Psi_A   \Psi_A^*  +    |c_B|^2   \Psi_B   \Psi_B^*  
+   c_A  c_B^*  \Psi_A  \Psi_B^*  +    c_B  c_A^*  \Psi_B   \Psi_A^*    \;,    \label{decohere1}
\eea
having  four peaks.      The nondiagonal components however quickly die out due to the scattering with the envronment particles  \cite{Joos1}-\cite{Zurek2}, typically as
\be   \rho({\bf r}, {\bf r}^{\prime}; t)=        \rho({\bf r}, {\bf r}^{\prime}; 0) \cdot   \begin{cases}
   \exp [ - \Lambda t  | {\bf r}- {\bf r}^{\prime}|^2 / 2 \lambda^2 ]  \;,    & \text{if}   \quad   | {\bf r}- {\bf r}^{\prime}|   \ll  \lambda\;,    \\
    \exp [ - \Lambda t ]\;,      & \text{if}    \quad    | {\bf r}- {\bf r}^{\prime}|   \gg  \lambda\;,  
\end{cases} \label{decohere2}
\ee
where $\lambda$ is the wavelength of the environment particles, and the decoherence rate
\be  \Lambda \equiv   n v   \sigma \;    \label{decohere3}
\ee
is given by the density $n$ times velocity  $v$   (flux) and $\sigma$ the cross section. Concrete decoherence rate and its dependence on the kind of environment and on the types of macroscopic bodies  are found in \cite{Joos1}-\cite{Zurek2}.

It is reasonable to assume that such a decoherence mechanism, applied to the experimental devices,  entangled with the environment (the rest of the world),  indeed  explains the classical state of the measurement-recording device after each measurement  -  the uniqueness of each experimental outcome -,  providing us with  one of the important ingredients for understanding the quantum measurement processes. 

However, there are reasons to believe that this is not the end of the story.  In author's view, it is necessary to consider the whole problem from a wider perspective of what may be called    ``great twin puzzles of physics today":   the general  quantum measurement problems  \cite{WheelerZ}-\cite{Peres}  on the one hand, and emergence of the classical mechanics from QM, on the other.   Even though these two classes of the problems are often discussed together, causing sometimes misunderstanding  rather than clarifying either of them,  they are mostly independent issues, and should be discussed independently. This is so, even if  there are a few key questions which link the two classes of the problems: the classical behavior of the measurement devices  mentioned above is one of them. Another is the concept of the center-of-mass (CM) position and momentum of a macroscopic body, which requires necessarily a measurement to define them as the initial condition for studying the time evolution of the system.

The first of the twin puzzle,   the quantum measurement problems, has a long history \cite{WheelerZ}-\cite{Peres}.  This is not the place to discuss or review them. Let us just note that these issues have been revisited recently in  \cite{KK, KKTalk} from a new perspective, according to which  the quantum fluctuations represented by the wave function are real physics, and 
 the expectation values of all dynamical variables (giving the relative frequencies of different experimental outcomes) 
 are to be regarded as the fundamental law of QM, replacing Born's  rule. The wave function collapse is just a way we perceive the
 unique outcome of each measurement: due to the pointike nature of the basic building blocks of our world (and of QM itself, see Sec.~\ref{Qratio} below),
 each quantum measurement is essentially spacetime-event-like  \cite{KK,KKTalk}.

It is the purpose of the present work to discuss the second of the twin puzzles:    emergence of the classical mechanics from QM.  As noted already, the idea of environment-induced decoherence can be thought as providing the basic understanding of this problem.  Strictly speaking, however, this idea is neither sufficient nor necessary to explain why  macroscopic bodies obey Newton's equations.  First,  none of the scattering processes  which are studied to calculate the decoherence rates \cite{Joos1} -\cite{Zurek2}  represent the main physical processes of interest in
proving the  simple law  (Newton's equations)   for the CM motion of a macroscopic body, e.g., under weak gravitational forces or external electromagnetic fields.  

 Secondly, very generally, any matter,  in any state,  is in a quantum mechanical state at temperatures sufficiently close to the absolute zero.     At $T=0$  any system is in its unique quantum ground state  (the 1st law of thermodynamics, or the Planck-Nernst law). 
As  is familiar from the study of various critical phenomena,  many systems, gas or liquid,  via a phase transition go into   various collective, macroscopic quantum mechanical states below some critical temperature \cite{Cornell} -\cite{He3}.   Superfluidity and superconductivity  are well-known examples; the ultra cold gas of atoms forming Bose-Einstein condensation is another.  On the other hand, the conducting electrons inside a metal cable  at room temperatures, or  light, a gas of free bosons (the photons) at any temperature, are macroscopic but quantum mechanical systems.  What these considerations tell us is that temperature enters critically in the discussion on ``how classical behavior arises from QM?",  but in a very nontrivial, system-dependent way.  
 It is necessary to formulate the problem  itself more carefully,  including, e.g., what is meant by  a ``macroscopic body". 

Thirdly,  the presence of a pre-existent environment is not always necessary. An example is provided  by the  experiment  by Arndt et.al.,  done by using  beams of  $C_{70}$ molecules  \cite{C70}.  $C_{70}$ molecules are bombarded by a laser light, get excited and then emit photons,\footnote{It is of course possible to interpret  absorption of laser light - excitation - and emission as a special type of scattering process, not to be considered as distinct from those considered in  \cite{Joos1} -\cite{Zurek2}. The point however is that any isolated macroscopic body at finite body temperatures 
emits photons and decoheres  (see Sec.~\ref{Decoherence} below).    The processes studied in   \cite{C70} may be regarded also as an example of this general phenomenon. }    before entering a Talbot-vonLau interferometer \cite{Clauser}. It is shown experimentally  that  the entanglement-decoherence, caused by the emission of photons which carry away information, is sufficient for $C_{70}$  particles to  behave ``classically", losing the ability of forming coherent superposition of states
(see more about it  in  Sec.~\ref{Qratio}  and  Sec.~\ref{QCL}).  A general theoretical analysis of   
 these ``self-decoherence"   processes  by the photon emissions by isolated,  highly excited large molecules  (or small particles),     was given by Hansen and Campbell   \cite{Hansen}.

Keeping all these points in mind,  we propose to study in this work  
the center-of-mass (CM)  motion of an isolated macroscopic body with  finite body temperatures  in the vacuum   \footnote{We 
assume that,  {\it    for the purpose of the discussion here}    the nontrivial quantum properties of the vacuum  (see, e.g., \cite{Birrell}), such as the Hawking radiation, the Unruh effects, the vacuum polarization and the Casimir effects,  etc.,  can be safely neglected.} (or,  in a cold environment, by using the terminology used  in \cite{Hansen}).  We mean by  a ``macroscopic body" any macroscopic composite state  (bound state) with a well-defined mass and size (a piece of solid, a dust particle, metal pieces, crystals, stones and rocks, biological bodies, planets, stars, etc.). In other words, we assume that these macroscopic bodies have well-defined and limited spatial supports of their internal wave function.  This will exclude a volume of free  atomic or molecular gas,  but may include  a droplet of liquid or a volume of ocean water, as long as they occupy well-defined volume in space.   

This work is  organized as follows.  In Sec.~\ref{basic}  we review three quantum mechanical aspects  in the present context, (i) Heisenberg's uncertainty relations for  the CM of a macroscopic body, (ii) absence of diffusion due to its large mass,  and (iii)  a finite body temperature, implying a radiating, metastable state. 

 These constitute  the necessary and sufficient conditions for a macroscopic body to have a well-defined  CM position and momentum at each instant of time, once they are determined experimentally with an arbitrary precision  (the point (i) above).  
 
Along the way,  the resolution of the old Born-Einstein  (not Bohr-Einstein!)  dispute is explained and  the key role  of the temperature (the missing piece of the puzzle) is illustrated. 
  
Once 
the concept of an arbitrary narrow  (in ${\bf P}$ and  in ${\bf R}$) wave packet for the CM of a macroscopic body  is established as something realizable by 
an experiment as the initial condition,  the derivation  of Newton's law  in  Sec.~\ref{Newton}   is, in principle,  a straightforward application of Ehrenfest's theorems;   the derivation actually involves several subtle features and is not trivial.  It is illustrated in the case of weak gravitational forces, a harmonic-oscillator potential, 
and weak external constant  electromagnetic potential, varying slowly in space. The derivation of  Hamilton's equations for many (macroscopic) - body 
problems is also discussed.   

In Sec.~\ref{QCL} the boundary between the quantum and classical physics  is discussed.  First we draw lessons from several experiments  (Sec.~\ref{temperature}),  made with the aim of illustrating quantum-mechanical nature of various mesoscopic-macroscopic bodies at sufficiently low temperatures. Though brief,  this discussion illustrates clearly 
how the QM-CL temperature boundaries depend on the system considered, and what their significance is.

 In Sec.~\ref{Qratio}   an approximate but simple universal criterion is proposed to tell whether a (microscopic or macroscopic) particle is classical or quantum mechanical,  according to the ``quantum ratio",     $R_q /L_0  \lesssim  1$ or $R_q/ L_0  \gg 1$,  
 where   $R_q$ is the quantum fluctuation range of its CM,   and  $L_0$ is the body's size.

Sec.~\ref{Evolution}  confronts the characteristics  of time evolutions in  classical and quantum mechanics.
This discussion somewhat unexpectedly leads us to a key observation on how the so-called quantum nonlocality (entanglement) 
and the dynamical causality  can coexist in quantum mechanics, without internal  contradictions. And this observation, in turn, shows why the so-called hidden-parameter models cannot replace quantum mechanics, without sacrificing  causality.

We summarize and conclude with some reflections in Sec.~\ref{Summary}.

\section{Classical trajectory for  the CM of a macroscopic body   \label{basic}}

We first review three aspects of QM in the context of  systems we are interested in:  an isolated macroscopic body in the vacuum, and with finite body temperatures.
\begin{description}
  \item[(i)]   Heisenberg's uncertainty relations for the CM of the macroscopic body;  
  \item[(ii)]   Large mass (large number of atoms and molecules composing it) implying absence of diffusion (for the CM);
  \item[(iii)]   The radiating,  mixed-state nature (decoherence) of the body.
\end{description}
Each of these, separately, is  well known.  
However, the
demonstration that these, taken together,  precisely  provide the (necessary and sufficient)  conditions for   the CM of a macroscopic body at finite temperatures  to  behave classically,
 is perhaps  new.

Note that this is so even if, in general, the microscopic degrees of freedom   (the electrons, atoms and  molecules)      composing the  macroscopic body under consideration (a macromolecule, a metal ball, a rock, planets and stars) will continue to behave quantum-mechanically.  Indeed, the body under study   is always a quantum-mechanical bound state, whose size and other properties are determined by the bound-state wave functions, even though at certain large scales onwards, an isolated body may well be described more simply as e.g., a classical gravitational bound state. 
 
The following discussions also clarify some conceptual issues, such as the Born-Einstein debate and its resolution, illustrating the central role of the temperature in the whole problem,  and shed light to the poorly-defined but familiar  ``puzzle" \cite{Joos}   why  microscopic particles are ``usually" in momentum eigenstates, whereas macroscopic ones are ``always" in a position eigenstate.

\subsection{Heisenberg's uncertainties    \label{Heisenberg} }

For a microscopic particle, an atom or molecule,   the derivation of Heisenberg's uncertainty relation is well known.
The canonical position and momentum operators of a particle,   $x, p$, satisfy 
\be   [x,p] =i\hbar\;.  
\label{commut} 
\ee
The uncertainty of $x$ or of  $p$  is defined as their dispersions 
\be \Delta x \equiv  \sqrt{\brc (x-x_0)^2 \ckt}, \qquad   \Delta p \equiv  \sqrt{\brc (p-p_0)^2 \ckt}\;,
\ee
where
\be   x_0 =\brc x \ckt =\brc \psi|x|\psi \ckt  \;, 
\quad  p_0 =\brc p \ckt =\brc \psi|p|\psi \ckt  \ee
are  their mean values  in the state $\psi$. 
The derivation of Heisenberg's uncertainty relation,
\be \Delta x \cdot  \Delta p  \ge  \frac{\hbar }{ 2}\;. \label{indeterm}\ee 
is found in any QM textbook. 
Extension to the case of a particle in three-dimensional space is straightforward. 

Let us now consider the case of a macroscopic body, made of  a large number $N$ of atoms and molecules, e.g.,  $N\simeq 10^{23}$,   $N\simeq 10^{57}$, etc. 
The center-of-mass position, the total mass and the total momentum  are  defined as
\be     {\bf R} =  \frac{1}{M}    \sum_{i=1}^N   m_i   {\bf r}_i\;;  \qquad  M=  \sum_{i=1}^N  m_i\;,\qquad    {\bf P}=  \sum_i {\bf p}_i\;, \qquad {\bf p}_i=- i \hbar \nabla_i
 \label{simple1}
\ee
It follows that   (${\mathbf R}= (X_1, X_2, X_3)$)
\be  [{X}_k, {P}_{\ell} ] =      \frac{1}{M}    \sum_{i=1}^N   m_i  \,   ( i\hbar   \,  \delta_{k \ell})   =   i\hbar    \,  \delta_{k \ell} \;, \qquad    k, \ell =1,2,3\;. \label{simple2}
\ee

For simplicity let us consider the total Hamiltonian of the form (see Appendix~\ref{coordinates}), 
\be  H= H_{CM} +   H_{int}\;,   
\label{TotHam}  \ee 
\be       H_{CM}=  \frac{{\bf P}^2}{2M} +  V({\mathbf R})\;;  
  \qquad     H_{int}= \sum_{i=1}^{N-1}    \frac{{\hat  p}_i^2}{2\mu_i} +   V_{int}({\hat  {\bf r}}_1,  {\hat  {\bf r}}_2, \ldots   {\hat  {\bf r}}_{N-1}) \;,
\label{Hamiltonians}  \ee
where  the  internal
 interactions  binding the body depend  only on the relative positions 
\be     V_{int}({\hat  {\bf r}}_1,  {\hat  {\bf r}}_2, \ldots   {\hat  {\bf r}}_{N-1})  \;.  \label{form} 
\ee
The internal (relative) coordinates  $ {\hat  {\bf r}}_i$  are defined by 
\be    {\bf r}_i = {\bf R} +   c\,   {\hat  {\bf r}}_i\;, \quad  i=1,2,\ldots, N\;; \qquad   
    \sum_{i=1}^N    m_i   {\hat  {\bf r}}_i = 0\;.    \label{relative} 
\ee
$c$ is a constant appropriately chosen so that the change of the coordinates has a unit Jacobian   (see (\ref{Jacobian}), (\ref{good})) and  $\mu_i$'s  are the reduced masses
(see (\ref{reducedmass}), (\ref{good})).
Our choice of the internal coordinates  and the consequent change of variables  from
 $ \{ {\bf r}_1,   {\bf r}_2,    \ldots     {\bf r}_N  \}  $  to $  \{ {\bf R},   {\hat {\bf  r}}_1, \ldots      {\hat {\bf  r}}_{N-1} \} $,
which will be frequently used 
below to separate the dynamics of the center of mass  from the internal motions of the body,          
is summarized in Appendix~\ref{coordinates}.

The  Hamiltonian of the form (\ref{TotHam}), (\ref{Hamiltonians}),    allows us to write the wave function in a factorized form, 
\be   \Psi({\bf r}_1,  {\bf r}_2, \ldots  {\bf r}_N) =   \Psi_{CM}({\bf R}) \psi({\hat  {\bf r}}_1,  {\hat  {\bf r}}_2, \ldots   {\hat  {\bf r}}_{N-1}) \;.   \label{factorization}
\ee
Repeating the same steps as for  a particle,  (\ref{commut}) - (\ref{indeterm}),     but  by using the wave function   $\Psi$   (\ref{factorization})    instead of  $\psi$,    
 one finds  straightforwardly from (\ref{simple2})  
\be \Delta X_k  \cdot  \Delta P_{\ell}   \ge  \frac{\hbar }{ 2}   \, \delta_{k \ell}     \;. \label{indetMacro}   \ee

Needless to say, we are using the nonrelativistic approximation here, for simplicity.
Take for instance the hydrogen atom.  The mass of a hydrogen atom in the ground state,  is not equal to  $m_e + m_P$, but
$m_H= m_e + m_P -  14 \,{\rm e\!V}/c^2 $,  by taking into account the binding energy.  However the error we make by   approximating the total mass by the sum of the electron and proton mass  is of the order of 
\be             \frac  {14 \, {\rm eV}/c^2 }{m_H} =    \frac  {14 \, {\rm eV}  }{0.94  \, {\rm GeV}}      \sim   O(10^{-8})\;.
\ee   
The same kind of errors will be even  smaller  in the case of the molecules, as their binding energies are in general smaller than the case of atoms.

Clearly this kind of argument cannot be applied to the level of the atomic nuclei,  where the binding energy is  relatively much larger.       
The problem might look even more serious if  a similar consideration is  extended to a more fundamental level.  It is known today that  the protons and neutrons which  make up the atomic nuclei are actually bound states made of the quarks and gluons,  and described by Quantum Chromodynamics  (QCD).  The mass of the nucleons, $\sim  940  \, {\rm MeV}/c^2\sim   1.67 \cdot 10^{-24}$  g,  
is mostly the effect of the  strong gauge interaction dynamics, with the contribution of the  up and down quark masses    ($m_u\sim  5 \, {\rm MeV}/c^2$, $m_d\sim  10 \, {\rm MeV}/c^2$)  being only a small portion of it.  

The problem is that the proton and neutron are  highly relativistic  quantum bound states,  for which the simple  classical  discussion such as (\ref{simple1}), (\ref{simple2}) cannot be  applied.  

Nevertheless,  nothing prevents  us from considering a macroscopic body as made up of  $N\simeq 10^{23}$ or    $N\simeq 10^{51}$    {\it   atoms or molecules},
which are in turn bound states of the electrons and the atomic nuclei (protons and neutrons),     and for the purpose of the discussion  here,  from disregarding  how the protons and neutrons 
are themselves made of  the quarks and gluons \footnote{Today  we know that  quantum chromodynamics  (QCD)  describes correctly the properties of the proton, neutron, and other strongly-interacting particles such as pions and $\rho$  mesons. }.  
 The understanding of how a macroscopic body is made out of atoms and molecules does not require - say, most of time -   the knowledge of how  the physics at smaller scales explain the properties of the atomic nuclei:    it is  a general character of physical laws.

With this understanding,  (\ref{indetMacro}) can be taken as  the correct Heisenberg's uncertainty relation for the center-of-mass position and momentum of a macroscopic body  \footnote{This fact is  often simply assumed.  It leads to the limits  in the resolution  in precision measurements involving macroscopic bodies and  
optical instruments,   known as  the standard quantum limit  (SQL)   \cite{SQL}. }. 

What is most remarkable is the fact that a macroscopic body, e.g., made of $N= 10^{23}$ or  $N= 10^{51}$     atoms and molecules, satisfies the same Heisenberg's uncertainty relations 
for its CM position and momentum,    as that for a single constituent atom. {\it  Heisenberg's uncertainties - the effects of quantum fluctuations -  do not pile up. }

But this means that  for an experimental  determination of the position and momentum of the CM of a macroscopic body, Heisenberg's uncertainty  constraints  may be regarded
unimportant.   
To have some concrete ideas,  one may take for instance, somewhat arbitrarily,    the  experimental uncertainties   $   \Delta X,  \Delta P$    of the order of 
\be     \Delta P \sim    10^{-2}  ({\rm g}) \times   \frac{10^{-4}   {\rm cm}}{\rm sec}  \sim   10^{-6}  \,  \frac  {{\rm g} \, {\rm cm} }{\rm sec}\;;   \qquad     \Delta X \sim  10^{-4} \, {\rm cm}\;.
\label{RuleofT}
\ee   
  Their product is   
\be   \Delta P \cdot \Delta X \sim  10^{-10}   {\rm erg} \,  {\rm sec}    \;,  
\ee
many orders of magnitude   larger than the  quantum mechanical lower bound,    $ \hbar  \sim 1.05  \times 10^{-27}   \,   {\rm erg} \, {\rm sec}.$
Thus even allowing for  much  better precision by several  orders of magnitudes than the rule-of-thumb (\ref{RuleofT}),   both for the position and for the momentum,  one can still regard
the Heisenberg uncertainty relations insignificant.  To sum up,  one can effectively take
\be  \Delta P\approx 0\;, \quad  \Delta X \approx 0\;  \label{Definite}
\ee
 for macroscopic bodies.

In other words,    the position and momentum  of (the CM of) a macroscopic body can be measured and determined  simultaneously
 with  an ``arbitrary" precision,  at  macroscopic scales.   These data serve as the initial condition for Newton's equation of motion.

\subsubsection{Detection of the gravitational waves}

An alert reader  might wonder how simple considerations as  (\ref{RuleofT})-(\ref{Definite})  can resist a scrutiny in view of the extraordinarily    precise detection of the gravitational waves  at LIGO and VIRGO
laboratories  made recently  \cite{GW150914}.  
In the first observation of GW150914  performed at the two LIGO detectors,   to fix the idea,   $\sim  40$ Kg  mirrors used in the Michelson-Morley interferometer   oscillate  about  $\pm 10^{-16}$ cm,  with frequency around $100$ Hz.
Thus  the momentum and position uncertainties of the mirrors can be taken to be  of the order of 
\be  \Delta P \sim  4\cdot 10  \cdot 10^3 \cdot  10^{-16} \cdot 100 \sim   4 \cdot 10^{-10} {{\rm g\, cm}}/{{\rm sec}}  \;, \qquad   \Delta X \sim 10^{-16} \, {\rm cm}\;, 
\ee
such that their product is 
\be      \Delta X \cdot  \Delta P \sim   4   \cdot   10^{-26}  \, {\rm erg}\,\, {\rm sec}  \;,  
\ee
a factor $\sim 10^2 $ above the quantum-mechanical  lower bound   (\ref{indetMacro}),   i.e.,     $\hbar /2  \sim 5 \cdot 10 ^{-28}$   erg sec.    
 See Ref. \cite{Braginsky} for a more careful analysis, where  it is shown that  the main  source of the  sensitivity limit for the gravitational wave detection  comes from the effects related to  the quantum fluctuations of the photons in the Fabry-Perot cavity, related to the radiation pressure and  laser shot noise, rather than the Heisenberg uncertainty of the mirror CM position and momentum.      In other words, 
  the observed gravitational-wave-induced oscillation of the mirrors  is a classical motion.

\subsection{Diffusion of the wave packet   \label{Diffusion}  }  

Another textbook result in QM  is  that  the wave function of a free particle
 (e.g., an electron, an atom)  diffuses in time.  For instance,
a  Gaussian wave packet of size $\sim a$   at time $t=0$,  
\be    \psi(x, 0) = \const  \, e^{- x^2 /  a^2}       \label{Gaussian}
\ee
evolves,  according to the Schr\"odinger equation,  
\be    i \hbar   \frac{d}{dt} \psi(x, t) =   H\,   \psi(x, t) \;,   \qquad  H =  \frac{ p^2}{2 m} \;
, \qquad  p= - i \hbar \frac{\de}{\de x} 
\ee
to a wider distribution, 
\be   |\psi(x, t)|  =   \const  \, e^{- x^2 /  a(t)^2} \;, \qquad     a(t) =  \sqrt{   a^2 +   \frac{4 \hbar^2 t^2}{m^2  a^2}  } \sim    \frac{ 2\hbar t }{m a}   \label{diffusion}
\ee
after time $t   (\gg  \tfrac{m a^2}{2 \hbar})$.   Note that the initial  momentum uncertainty 
\be      \Delta p \sim \frac{\hbar }{\Delta x} \sim \frac{\hbar }{a} 
\ee
is responsible for the spreading of the wave packet.

{\it   The mass plays an essential  role in the  diffusion of a  (free)  wave packet. }  To have a concrete idea,   
the time required for doubling the wave packet size from $1  \mu$ to  $2  \mu$   due to the evolution  (\ref{diffusion}) 
is compared  for several types of  particles, in Table~\ref{diffusiontime}.

 \begin{table}
  \centering 
  \begin{tabular}{|c|c|c|  }
\hline
 particle   &   mass  (in $g$)  &    diffusion time   (in $s$)  \\   \hline
 electron   &   $9  \cdot  10 ^{-28}  $  &     $10^{-8}   $   \\
   hydrogen atom   &   $1.6  \cdot  10 ^{-24}  $    &    $1.6 \cdot 10^{-5}  $  \\
   $C_{70}$ fullerene  &   $8   \cdot  10 ^{-22} $  &     $8 \cdot 10^{-3}$    \\
   a stone of $1g$    &     $  1  $    &     $10^{19}      $  \\  
\hline
\end{tabular}
  \caption{ \footnotesize  Diffusion of the free wave packet for different particles. For definiteness we assume the initial wave packet size of $1 \mu$;  
  the diffusion time is defined as $\Delta t$ needed for doubling its size.  For a  macroscopic particle of $1 g$, the doubling time,  $   10^{19}  {\rm sec}   \sim 10^{11}  {\rm yrs}$,
   exceeds  the age of the universe.
       }\label{diffusiontime}
\end{table}

The conclusion is that  the diffusion of a  free wave packet - quantum fluctuations - of the center-of-mass position is negligible  for  a macroscopic object,  due to its large mass.   This difference explains  and resolves  an apparent (and  loosely defined) puzzle often discussed  in the literature \cite{Joos},  why  microscopic particles  (e.g., electrons) are ``usually"  in momentum eigenstates,  whereas a macroscopic body is
``always"  in a position eigenstate. 

The above simple argument also captures the qualitative difference  in  the role  quantum diffusion plays in the case of  microscopic and macroscopic bodies.   
 This being so,  however,  some caution and clarification are necessary.

  First of all,  the diffusion of the wave function reviewed above refers to a free particle.    
The atoms, molecules  and  electrons composing a macroscopic body might remain  quantum mechanical,  but their wave functions do not diffuse.  They are described by stationary bound-state wave functions. The size  $L_0$   of the  body   (a macro molecule, a crystal ball,    a 
piece of a stone, the earth, the sun,  etc.)  is determined by the dynamics which bind the constituent electrons, atoms  and molecules  in 
a bound state, the lattice structure,  etc.,   in other words  by the extension (i.e., spatial support)   of the wave functions   describing the bound state.

It is essential that the  spread of its CM wave packet, $\Delta_{CM}$,  is  distinguished from the size  $L_0$ of the body.  
Their ratio (a particular example of the quantum ratio,  discussed later, in Sec.~\ref{Qratio}),   determines if the body behaves classically or quantum mechanically.

    An experimentalist  capable of  determining  (measuring)  the size of the body  $\sim L_0$ will certainly be able to measure  its CM position  with precision,  
\be     \Delta_{CM}    \ll   L_0\;;      \label{precision}    
\ee
but it does not mean that  such a measurement has indeed been made.   A macroscopic body  may well be  in a state described by a  CM  wave packet such that   
\be     \Delta_{CM}    \ge   L_0\;, \qquad {\rm or \,\, even}  \qquad    \Delta_{CM}    \gg   L_0\;.   \label{possible}      
\ee
The point is that  the argument based on the particular, Gaussian wave packet,  (\ref{Gaussian}),  presumes that the position
and  momentum measurements had been done previously, with the precision  
    compatible with  the minimum uncertainty product,  $\Delta x \Delta p \sim \hbar /2$.  
  That would tacitly imply a state described by the inequality,   (\ref{precision}). 
  
  In general,  however,  a macroscopic  (and for that matter, also a microscopic) body may well be in the state,   (\ref{possible}),  i.e.,   in a quantum state with    $\Delta X \Delta P  \gg  \hbar$, i.e.,   far from the one with the minimum uncertainty.

\begin{figure}
\begin{center}
\includegraphics[width=4in]{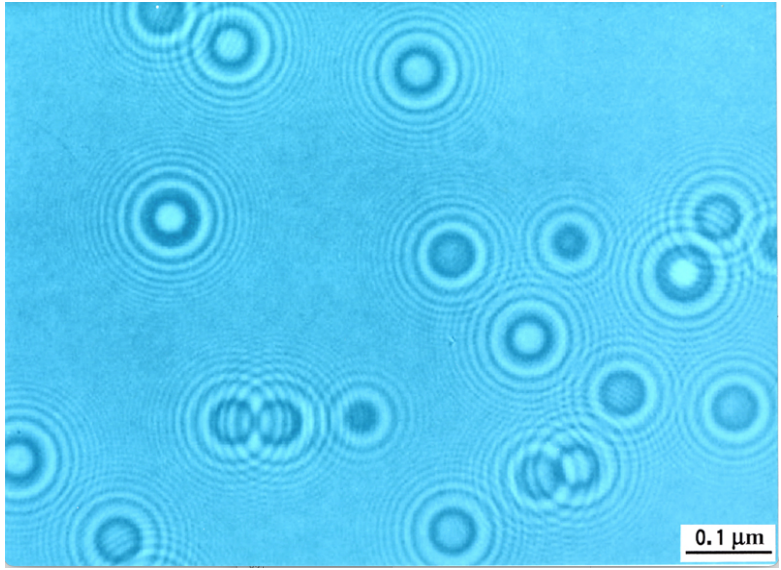}
\caption{\small   Diffusion, diffraction and interferences of the electron waves.   They have been produced by letting a $50$ kV  electron beam
fall on a thin collodion (celluloid) film.  Courtesy by A. Tonomura.   }
\label{Tonomurawaves}
\end{center}
\end{figure}

For microscopic particles,  
 if a (transverse) position measurement is done by 
impinging the beam on  a film with a hole  of diameter $d$,  it receives a 
momentum uncertainty  \cite{Heisenberg}, 
\be  \Delta x \sim  d\;; \qquad .^.. \quad    \Delta p \ge   \frac{\hbar}{ d} \;.  
\ee
so that it diffuses,  and if there are many holes in the sheet,  the particle waves  from different holes interfere. 
 The  image  of the electron wave ripples (diffraction and interference),  shown in Fig.~\ref{Tonomurawaves},  
 which look identical to  water ripples  on a puddle surface  produced  by rain drops,  but are smaller by a factor 
$\sim 10^{-6}$ in scale, have been obtained by A. Tonomura.
As the  law of diffusion of a free particle wave is known,  an initial wave packet of an arbitrary size can be  experimentally prepared, and used as the initial condition. 

For macroscopic bodies, it is important to distinguish the two cases   (\ref{precision})  and (\ref{possible}).    In the second (quantum)  case where the size of the body  $L_0$  is smaller than the extension of the wave packet of the CM of the body,  the discussion made above for microscopic particles apply equally well to macroscopic bodies.   (See Sec.~\ref{Qratio}).
This fact is especially relevant in the consideration of macroscopic, quantum mechanical systems  realizable at extraordinarily low temperatures.

\begin{figure}[h]
\begin{center}
\includegraphics[width=4.5in]{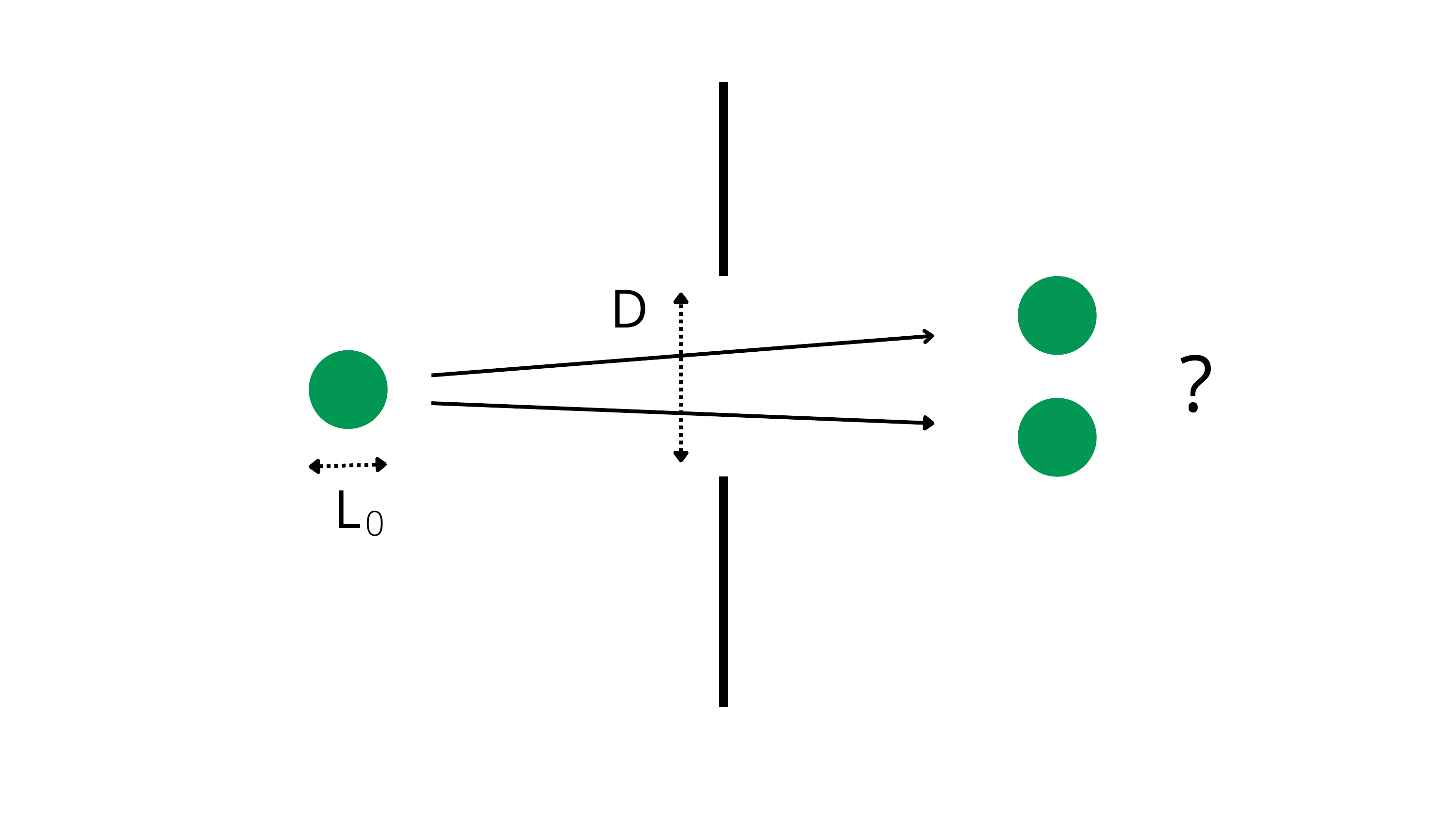}
\caption{\small  The passage of a macroscopic particle of size $L_0 $  and with a well-defined CM position,   $ \Delta_{CM}$,   $L_0 \gg  \Delta_{CM}$,     through a hole of diameter  $D> L_0$.  The passage does not increase momentum uncertainty:
there is no diffraction.   }
\label{Single}
\end{center}
\end{figure}

In the case (\ref{precision}),  the situation is drastically different.  A  (transverse) position determination 
of a macroscopic particle  requires a hole of diameter $D$ larger than its size, $ L_0$ (otherwise the body will not pass).  If it passes its transverse position will have been measured with the precision,    
   \be  \Delta X =  D   >   L_0, \ee   
 but it is simply  due to the lack of information which  path the body has taken inside the hole  (Fig.~\ref{Single}).  
The (transverse)  momentum uncertainty caused by  such a position measurement 
\be    \Delta P \sim  \frac{\hbar}{D}   <    \frac{\hbar}{L_0} \;,\label{wouldbe}
\ee
 is much smaller than the  original momentum uncertainty of the  CM   wave packet,
\be      \Delta P_{CM}  \sim     \frac{\hbar}{\Delta_{CM}}   ( \gg      \frac{\hbar}{L_0} ) \;.\label{momunc}
\ee
In other words, the  position measurement in this case  does not affect the original momentum uncertainty, in contrast to the case of a microscopic particle. The passage through a hole does not induce a diffraction/diffusion of the wave packet.

\subsection{Decoherence:  a macroscopic body at finite temperatures   \label{Decoherence}}  

The third  condition for a macroscopic body to behave classically,  is decoherence. It is here that the finite body-temperature  plays an essential role,  which makes the body   
 in the vacuum under consideration a metastable, radiating,  mixed state.

The suppression of the diffusion of the CM of a macroscopic body
 was  regarded  by Born  as a sufficient reason for such an object to behave classically.  
  Einstein disputed it   by observing  \footnote{This debate is reported, e.g., in  \cite{Joos}.}  that a  superposition of the state
\be   |\Psi \ckt   = c_1  |  \Psi_1 \ckt  +   c_2   | \Psi_2 \ckt\;, \label{nopure}
\ee
where $\Psi_1$ and $\Psi_2$ are two narrow wave packets  centered at  ${\bf R}={\bf R}_1$ and at ${\bf R}={\bf R}_2$, separated by a macroscopic distance,  {\it is}  allowed by QM, i.e., as a solution of  the Schr\"odinger equation.  A state of this kind  certainly contradicts the notion of a classical particle, with its unique trajectory,  at a definite position at each  instant of time.

One way to explain away  this apparent paradox is to note simply  that a macroscopic body, {\it   at nonzero temperature},   is 
   a mixed state:  it is, at best, a metastable state,   
 \be  |  \Psi_1 \ckt  =    |  \Psi_1^{(0)}  \ckt   +    |  \Psi_1^{(1)}, \gamma  \ckt   +   |  \Psi_1^{(2)}, \gamma  \gamma  \ckt + \ldots   \label{macrobody}
 \ee
 where  $ |\Psi_1^{(0)}  \ckt $   is the effective  "wave function" of the metastable state describing  systems with atoms and molecules in excited states;   
 $  |  \Psi_1^{(1)}, \gamma  \ckt  $ with one emitted photon flying away, with the rest of the system still containing atoms and molecules
 in excited states,    $  |  \Psi_1^{(2)}, \gamma  \gamma  \ckt  $ with two emitted photons,  etc.   Similarly for  $ |  \Psi_2 \ckt$,
  \be  |  \Psi_2 \ckt  =    |  \Psi_2^{(0)}  \ckt   +    |  \Psi_2^{(1)}, \gamma  \ckt   +   |  \Psi_2^{(2)}, \gamma  \gamma  \ckt + \ldots   \label{macrobody2}
 \ee
  The coefficients 
  of different terms  in (\ref{macrobody}) and in  (\ref{macrobody2})   are omitted, as the norm of each term  depends on time anyway.    In the case of a single atom in a metastable, excited state, it is still possible to write a reasonably precise expression for    its (effective) wave function (see for instance \cite{KKGP});     in the case of a macroscopic body at finite temperature, with  many atoms and molecules in the metastable, excited states, such a description is neither available, nor really needed. The only thing needed is the fact that each of  the ``states",  
  $ |  \Psi_1 \ckt  $ and  $ |  \Psi_2 \ckt  $, is actually a mixed state,  so that their  quantum interference  is  negligible.      The density matrix  corresponding to  $\Psi$ of  (\ref{nopure}) in the coordinate representation is diagonal, in good approximation.

It is interesting to  compare the situation with the wave function of an $Ag$ atom, in the Stern-Gerlach experiment,  
\be     |\psi \ckt   = c_1  |  \psi_{1}  \ckt  +   c_2   | \psi_{2}  \ckt\;, \qquad    \label{SGatom}
\ee
where  $\psi_1$ and   $\psi_2$  represent the spin up and down states, respectively.  In an inhomogeneous magnetic field  in the ${\hat z}$ direction,    the two components 
$\psi_{1}$ and $\psi_{2}$  get separated in the vertical direction,  as the atom proceeds towards (say) the ${\hat x}$ direction.    
The two wave packets might well get  separated  by a macroscopic distance, before  the atom impinges upon the photographic plate. 
The wave function  (\ref{SGatom})  still represents a pure quantum state, and  the phase coherence  between $\psi_{1}$ and $\psi_{2}$  can be  
verified and measured, if    the two beams are reconverged  by another magnetic field   before they hit the photographic plate, in  
 an experimental setup known as the quantum eraser.

The contrast  between the macroscopic ``wave packets"  (\ref{nopure}) and the quantum-mechanical pure state  (\ref{SGatom}) 
  may be made manifest,    by assuming that  such a superposition  has been somehow prepared  at time $t=0$, and by  studying the time evolution  of the density matrix in the position representation.    The ideas of  decoherence   \cite{Joos1}-\cite{Arndt1}  is that because of the scattering - entanglement of the body  with the environment particles, nondiagonal elements of the density matrix  get rapidly damped in time, leaving the diagonal components, 
see  (\ref{decohere1})  -  (\ref{decohere3}),   
  showing that the 
system gets quickly converted to a mixed state.

The main observation in the present work is, however,  that  it is  sufficient that the body radiates, i.e., the body temperature is nonzero,  for it to be a mixture, and  to decohere. There is no need for a pre-existent environment.    We have already cited  in Introduction  the work  by Hansen and Campbell  \cite{Hansen}
which analyzes this sort of process for isolated large molecules, and the experiments by
 Arndt et.al. \cite{C70}  in which  $C_{70}$  beam, heated by bombarding laser beams, get decohered via emission of the infrared photons.  
See Sec.~\ref{QCL}  for more comments.

Some idea of  the decoherence rate for an isolated macroscopic body at finite temperatures comes from the following consideration.  We  first note that,  in the environment-induced decoherence studies \cite{Joos1}-\cite{Arndt1},     the decoherence rate calculated (\ref{decohere2})  tells essentially that the body under consideration decoheres when it  suffers from {\it any} scattering from the environment particles.
The expression (\ref{decohere3})  is  nothing but the total probability per unit time the body gets scattered, and its information (about the phase, the position, the angular momentum, etc.)  carried away and lost. 

An atom in excited state  or a metastable nucleus,   may be described by an effective wave function, having an energy level with a small imaginary part,  
\be   {\hat  \psi}_X(t)  =     e^{- i E t / \hbar}  {\hat  \psi_X(0)  } =   e^{- i E_R t / \hbar}   e^{- \Gamma t /2 \hbar }   {\hat  \psi_X(0) }  \;,      \label{smooth}
\ee
where $\Gamma$ is the total width,  or the total decay probability per unit time.  ($\tau = 1/\Gamma$  is the mean lifetime of the metastable atom or nucleus).  Such a wave function means that at time $t$ the probability that  the atom (or nucleus) in an excited or metastable state  has not decayed,  is given by \footnote{This is a large time approximation.  At times shorter than the inverse of the characteristic frequency of the system,    $P^{0}(t)$ actually behaves as   $P^{0}(t) \sim 1 - c \, t^2 + \ldots $, leading to interesting physical effects such as quantum Zeno effects.  We do not touch upon these problems here.  
}
\be    P^{0}(t)  \sim   e^{- \Gamma t}\;. 
\ee
$ P^{0}(t) $ significantly smaller than unity  (that is, $t > 1/\Gamma$)  means that the system is in a mixed state.   

For a metastable nucleus which decays via a spontaneous $\alpha$  emission, 
\be       {}^A_N \left( X  \right)    \Longrightarrow       {}^{A-4}_{N-2} \left( Y  \right)  + \alpha\;,
\ee
where $N$ is the atomic number and $A$ is the mass number,    it is tempting to write a wave function of the form, 
\be        | \psi \ckt  =     | X \ckt    +    | Y \ckt  | \alpha  \ckt\;  \label{nucleus}  \ee
 (ignoring  the coefficients in front of  the two terms: 
their norms depend on time due to the decay.)
Now the crucial observation is that an expression such as (\ref{nucleus})  represents a mixed state, not a pure state.  Coupled with possible ``alive" or ``dead" cat states, regarding the system 
\be       | X \ckt   \otimes |alive \ckt  +    | Y \ckt  | \alpha  \ckt    \otimes |dead \ckt \;,  \label{cats}  
\ee
as a coherent superposition of states leads to the notorious Schr\"odinger conundrum.    Actually,  interpreting correctly  expressions such as (\ref{nucleus}) or  (\ref{cats}) as  mixed states (at any time, the nucleus is either decayed or not decayed) and not  pure quantum states, 
  is all that is needed  \cite{KK,KKTalk} to eliminate altogether the familiar Schr\"odinger puzzle.

For a macroscopic body at finite temperatures the main process of relevance is the spontaneous emission of infrared photons from the atoms or molecules
in excited states.  The total  decay width of a single excited atom or molecule $\Gamma_i$  depends on the type of the atoms or molecules, the type of the excitations and on the excitation levels. Furthermore, the total decay rate of the body   depends on its  temperature ($T$)   (which determines the average excitation energy), and the number $N_0$  of the atoms and molecules in excited states.  The probability that the body has not decayed during the interval $t$
will have the form, 
\be   R \sim      \exp  [ - \sum_i^{N_0(T)}  \Gamma_i(T)   t ] =  \exp [-  \Lambda(T)  t ]  \;, \label{DecoMacro}
\ee 
where   $\Lambda(T) $ may be interpreted as the decoherence rate for an isolated body at body temperature $T$.  For   a typical macroscopic body at finite temperature,  
\be        T  \gg      T_0\;,    \qquad  T_0  =     \frac{h \nu }{k_B}  =      \nu   \cdot  4.8   \cdot  10^{-11}   \, K  \label{MacroT}
\ee
where $ h  \nu$ is the characteristic, lowest excitation energy of the system,     and with
$     N_0(T)   \gg 1\;,  
$
 the decoherence will occur basically instantaneously.     Just to have a quantitative  idea let us take for a small stone of $1\, g$  \footnote{One might want to use    in (\ref{estimate})
 a typical phonon excitation energy or their decay width, several orders of magnitude smaller than the characteristic atomic or molecular excitation energies,
  but the 
 essence of the argument remains unchanged.   }
 \be    \Lambda   \sim \sum_1^{N_0}     O(eV)    \sim    10^{23}   \,   eV \;.       \label{estimate} 
 \ee 
  {\it  Assuming} that  a coherent superposition of the form  (\ref{nopure}) has been, somehow,   prepared for it at time $t=0$,
 it decoheres in
 \be     \Delta t \sim    \frac  {\hbar}{  \Lambda }  \sim  10^{-34} \, s\;. 
 \ee

\subsubsection{Split wave packets for a   (finite $T$) macroscopic body:  an impossible notion}

Actually,   there is a better,  logically more consistent, way  to argue  why a pure state of the form (\ref{nopure}) does not occur in nature, for a macroscopic body at finite   temperatures.   The answer is,  as it follows from the  discussions of Sec.~\ref{Heisenberg} $\sim$  Sec.~\ref{Diffusion},   that  {\it    it is not possible to prepare  a doubly (or multiply)  split wave packet  such as  (\ref{nopure}),  experimentally. }

Let us note that both in quantum and classical physics, the initial condition to be used as the input of the dynamical evolution  of the system, is the result of a precedent measurement.  And it is here that   quantum and classical mechanics differ significantly.

In the case of microscopic particles,  a well collimated  beam of the electrons or atoms  (a wave packet of narrow width, i.e.,  with a well-defined transverse position)  may be  sent to a double slit, to yield a wave function of the form, (\ref{SGatom}), after the passage.  Famous experiments such as  \cite{Tonomura}  showing beautiful interference fringes \`a la Young,  demonstrate clearly  the pure-state nature of  $|\psi\ckt$.    
 Alternatively,   if the particle carries a spin, an initially well-defined wave packet can be  split in  two or more  (two if  $s=\tfrac{1}{2}$, three if  $s=1$, etc.)  components  in  an inhomogeneous magnetic field,  by using the familiar Stern-Gerlach experimental set-up.  
Either way, there is no difficulty in preparing experimentally a pure state having the   split-wave-packet form (\ref{SGatom}); it is a perfectly 
good new  initial condition.

 \begin{figure}
\begin{center}
\includegraphics[width=5in]{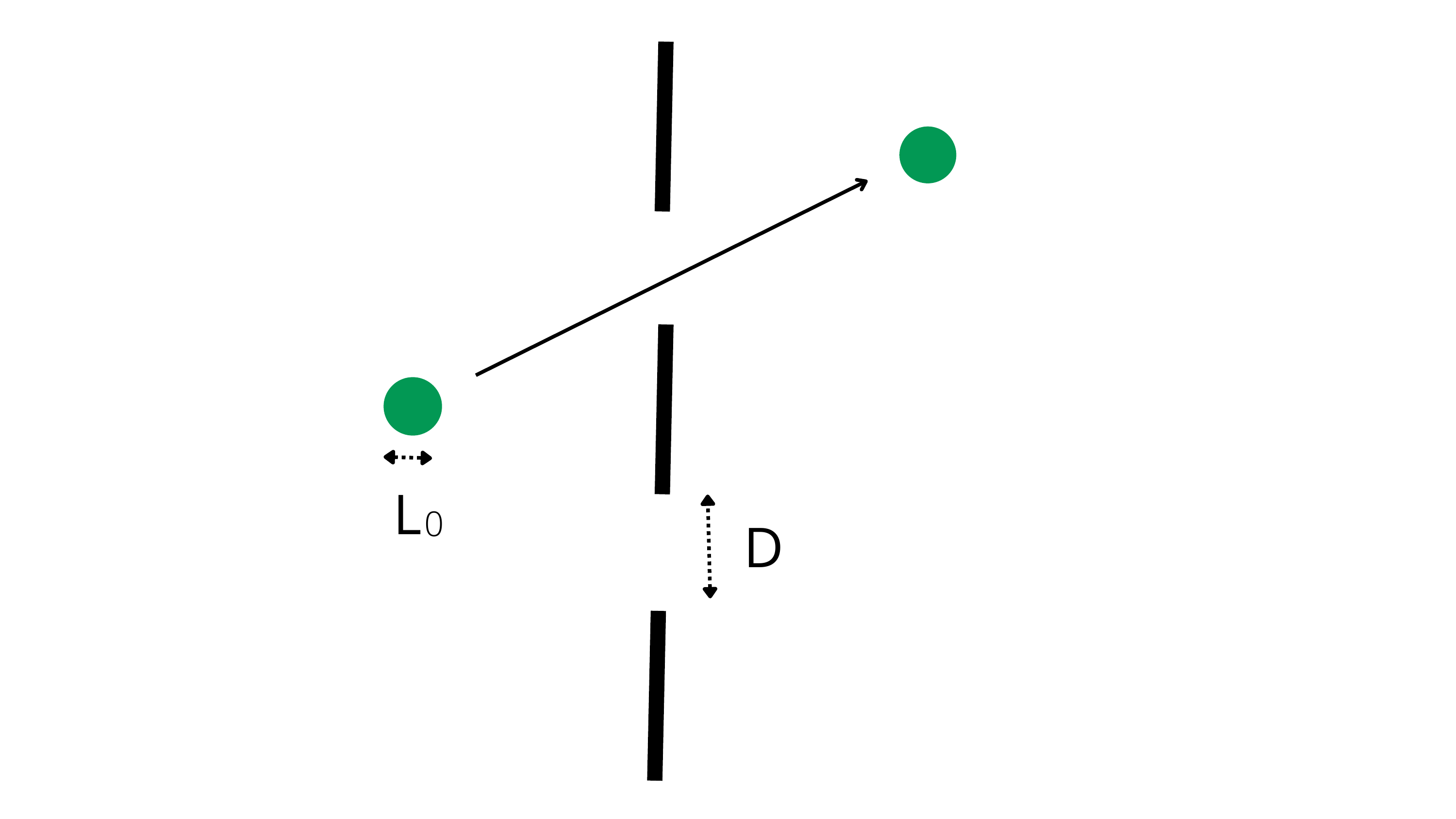}
\caption{\small  The passage of a macroscopic particle of size $L_0$,   through a  double slit, each hole with width  $D> L_0$.  The passage 
will leave the particle behind one of the slits, each time.}
\label{Double}
\end{center}
\end{figure}

On the contrary,  for a macroscopic body at finite temperatures,   
whose position and momentum have been both previously determined experimentally with an arbitrary precision,  (\ref{Definite})  and (\ref{precision}),   
there is no way to convert it to a coherent superposition of the form,
(\ref{nopure}). 
 A double slit  wide enough  for a macroscopic body to pass, will simply leave the particle either behind the one  slit or  the other,  as
 \be   \Delta_{CM} \ll L_0    \leq   D\;, 
 \ee
see Fig.~\ref{Double}.  It may pass through one or the other if the experiment is repeated \footnote{Obviously, here the original transverse  momentum  uncertainty is assumed to be such that one cannot predict 
which path the particle takes. 
}.  It would result in a statistical mixture, not a pure state \footnote{This indeed seems to be what happens to the $C_{70}$ molecules when they get heated up sufficiently by a raised laser power,  in the experiment of \cite{C70}.}.

The Stern-Gerlach  set-up  would  not help  either, to ``prepare"  a  doubly (or multiply) split wave packet for such a macroscopic body,   (\ref{precision}).   As it has a well-defined  position and momentum for its center 
of mass at any time,    it would simply follow a well-defined classical  trajectory  in a magnetic field  (and not  up {\it  or}  down),    
 depending on the orientation of its magnetic moment and on the gradient of the  magnetic field. This is what we know from classical physics, what is known as an empirical fact, and what indeed follows from QM,  see 
below,  (\ref{NoSG}).

 \subsubsection{Settling the Born - Einstein dispute}
 
The discussion above on the impossibility of a split wave packet for a macroscopic body,    concerns  radiating, mixed-state, thus classical,  body, at finite temperatures.    As recalled  in Introduction,  however,  
at temperatures sufficiently close to the absolute zero, any matter,  microscopic or macroscopic,  is quantum mechanical.  A macroscopic 
 coherent superposition (\ref{nopure})   (a pure state of this form)  is certainly possible, for such systems. 
  The pair of the SQUID states with macroscopic fluxes of opposite signs, with possible tunnelling between them \cite{Leggett},  are  a perfect analogue  of the split wave packet
   (\ref{nopure}).  
   
 We  conclude that Born and Einstein were both right in a sense,  i.e.,    each under different conditions.  At finite temperatures   a macroscopic body is a mixed state and the absence of the diffusion
 (due to their large mass)  is indeed the basic reason for it to possess a classical trajectory;   at the same time,  coherent superposition of macroscopically distinct states is simply  not possible.   On the other hand,  at temperatures sufficiently close to absolute zero,  any system is quantum mechanical, and a macroscopic,  a split-wave-packet state (\ref{nopure})  is  certainly possible.    Such a state might look counter-intuitive from our daily experiences, nonetheless is nothing special in QM -  in principle.   The only problem is that, {\it   for a macroscopic body},  it requires extraordinarily low temperatures (to avoid self-decoherence) and a good vacuum (to eliminate environment-induced decoherence). 
  Attempts to realize such a state experimentally  are  in \cite{County} -\cite{Brand}.  See Sec.~\ref{QCL} for further comments. 
 
  The temperature (and more generally, decoherence)  was the missing piece of the apparent puzzle - the Born-Einstein dispute.

\section{Derivation of Newton's equations  \label{Newton}}  

Once the concept of a dispersionless center-of-mass wave packet of a macroscopic body is  established, with the initial condition $({\mathbf R}_{CM}, 
{\mathbf P}_{CM})$  as a well-defined and experimentally realizable input   (Sec.~\ref{Heisenberg}  - Sec.~\ref{Decoherence}),   
 the derivation of Newton's equation  itself is  a straightforward (in principle)  application of Ehrenfest's theorem.

The Ehrenfest theorem for a  microscopic particle  (electron, atom, molecule)  reads
\be   \frac{d}{dt}  \brc \psi   |  {\bf r}  |\psi \ckt  =   \frac{1} {i\hbar}   \brc \psi   |  [{\bf r}, H]  |\psi \ckt =   \brc \psi  | \frac{\bf p}{m}  |\psi \ckt  \;; \label{Ehr1}
\ee
\be   \frac{d}{dt}  \brc \psi   |  {\bf p}  |\psi \ckt  =   \frac{1} {i\hbar}   \brc \psi   |  [{\bf p}, H]  |\psi \ckt =   - \brc \psi  |   \nabla V |\psi \ckt  \;.  \label{Ehr2}
\ee
If   $ |\psi \ckt $ represents a wave packet of more or less well-defined position and momentum,     
this looks almost like  Newton's equation.  As is well known, a straightforward interpretation of  the wave packet as something representing a  sort of matter distribution inside a classical particle,  and the conclusion that  (\ref{Ehr1}) and (\ref{Ehr2})   
constitute the derivation of Newton's equation in QM,   
 however,  
are not tenable.  For instance, the wave packet of an electron hitting a  potential barrier easily breaks into two, one transmitted
and the other reflected;  on the other hand, the electron itself never breaks up.   A classical particle does not
 break either:   it gets either transmitted or reflected, depending on its energy and the barrier height.

These observations capture the essence of how classical mechanics can emerge from quantum mechanics on the one hand, and at the same time
explains the difference in the behavior of a macroscopic body at finite temperatures  from what is expected from a  (typically) microscopic, quantum-mechanical particle.  
As discussed in Sec.~\ref{Qratio},   what distinguishes the latter  is {\it   the ability of being in a pure state described by  a wave function having a nontrivial structures,   such as split-wave packets or  phase and magnitude undulations typical of interference effects, 
or simply entanglement with other particles,      over distances larger than its size. }

\subsection{Ehrenfest's  theorem for  the CM of  a macroscopic body   \label{EhrMac}}

Let us now generalize (\ref{Ehr1}) and  (\ref{Ehr2}) 
to  a    multiparticle bound state described by the wave function, 
\be  \psi({\bf r})   \to    \Psi({\bf r}_1,  {\bf r}_2, \ldots  {\bf r}_N)\;,
\ee
where $N$ is some large, macroscopic number, typically  $N\sim 10^{23}$ for a  piece of material of  $\sim1$g,    $N \sim 10^{51}$ for the earth.
The CM position, momentum  and the masses are given by
\be     {\mathbf R} =  \frac{1}{M}    \sum_{i=1}^N   m_i   {\mathbf r}_i\;;\qquad    {\mathbf P}= \sum_{i=1}^N  {\bf p}_i\;, \qquad   M=  \sum_{i=1}^N  m_i\;,\label{CMposition}
\ee
The Hamiltonian is 
\be    H \to  H=      \sum_i    \frac{{\bf p}_i^2}{2m_i}  +  V ({\bf r}_1,  {\bf r}_2, \ldots  {\bf r}_N)\;.  \label{macro}
\ee

All nuclei, atoms, molecules, crystals and condensed matter  states   are described by the Hamiltonian of this form.    We assume that also a general macroscopic body under consideration (a metal piece, a dust particle, a stone,  a rock, planets and stars, galaxies)   is also  described by 
the Hamiltonian,   (\ref{macro}).
    For a neutral atom  (Comlombic bound states, with a definite size of the order of Bohr's radius,    $r_B \sim 10^{-8} cm$),  all interactions are internal interactions, the Coulomb interactions between the electrons and the nucleus, and pairwise  interactions among the  electrons.
          For simplicity,   we neglect writing explicitly  relativistic corrections and   spin-dependent terms in   $H$.

In the study of the atomic spectra  (the internal excitation levels),   one usually neglects the  dependence on the center of mass ${\bf R}$ in the potential, and the atom as a whole is assumed to move as a free particle.    A similar comment can be made about the nuclear and molecular spectra: the main interest is in the internal excitation  modes and energy spectra.

Here,    we are  interested in the motion of the center of mass,  and  the forces the composite body  feels as a whole.   Our choice of the coordinate system, which is convenient in separating the dynamics of the CM from the internal motions of the body,  
 which is used throughtout this work,  is summarized in Appendix~\ref{coordinates}.

The first  Ehrenfest's theorem  (\ref{Ehr1}) easily  generalizes to the multiparticle composite body. 
As
\be     \frac{d}{dt}  \brc \Psi   |  {\bf R}  |\Psi \ckt  =  \sum_i  \frac{d}{dt}  \brc \Psi   |  \frac{ m_i {\bf r}_i}{M}    |\Psi \ckt  =      \frac{1} {i\hbar}  \sum_i     \brc \Psi   |    \frac{ m_i}{M}      [{\bf r}_i, H]  |\Psi \ckt =      \brc \Psi   |    \frac{1}{M}  \sum_i   {\bf p}_i |\Psi \ckt \;,
\ee 
one finds 
\be     \frac{d}{dt}  \brc \Psi   |  {\bf R}  |\Psi \ckt  =  \brc \Psi  | \frac{\bf P}{M}  |\Psi \ckt  \;; \qquad  {\bf P}= \sum_i  {\bf p}_i \;.\label{New1}
\ee
If we define  the classical CM  position and momentum of the body as 
\be      {\bf R}_{CM}(t)  \equiv     \brc \Psi   |  {\bf R}  |\Psi \ckt \;, \qquad   {\bf P}_{CM}(t)   \equiv     \brc \Psi   |  {\bf P}  |\Psi \ckt \;.      \label{PosMom}  
\ee
then (\ref{New1})  reads 
\be      \frac{d}{dt}   {\bf R}_{CM}(t)   =    \frac {{\bf P}_{CM}(t)}{M} \;,       \label{New11}
\ee 
which  is the familiar relation between the position and momentum for a classical point particle.

To consider  the  second Ehrenfest equation  (\ref{Ehr2}), let us write the potential as 
\be    V (  {\bf r}_1,  {\bf r}_2, \ldots  {\bf r}_N) =     V (  {\hat  {\bf r}}_1,  {\hat  {\bf r}}_2, \ldots   {\hat  {\bf r}}_{N-1},   {\bf R})\;\ee
by introducing the relative (internal) coordinates ${\hat   {\bf r}}_i$   (Appendix~\ref{coordinates}) 
\be    {\bf r}_1 = {\bf R} + c\, {\hat  {\bf r}}_1\;; \quad  {\bf r}_2 = {\bf R} +  c\, {\hat  {\bf r}}_2\;; \quad \ldots,   \quad    {\bf r}_N = {\bf R} + c\,  {\hat  {\bf r}}_N\;;    \label{invert} 
\ee
\be     \sum_i  m_i   {\hat  {\bf r}}_i = 0\;. \label{relative2} 
\ee
Any $N-1$ of   $ {\hat  {\bf r}}_i $'s can be taken to be linearly  independent, e.g.,    $ {\hat  {\bf r}}_1,  {\hat  {\bf r}}_2, \ldots   {\hat  {\bf r}}_{N-1}. $
This choice leads to (see (\ref{Jacobian}),  (\ref{good}))
\be  c= \left(\frac {m_N}{M}\right)^{\tfrac{1}{N-1}}  \simeq  1 \;, \qquad  N \gg  1\;. \ee

We need to compute
\be     [ {\bf P}, H]  =\sum_i  [ {\bf p}_i, H]  =   -i \hbar \sum_i    \nabla_i   
\ee
Inverting  (\ref{invert}),  one has
\be         {\bf R} =  \frac{1}{M}    \sum_i    m_i   {\bf r}_i\;;\quad    {\hat  {\bf r}}_1 = \frac{1}{c} ( {\bf r}_1   - {\bf R})  \;; \quad \ldots ;    \quad   {\hat  {\bf r}}_{N-1} =   \frac{1}{c} (   {\bf r}_{N-1}  - {\bf R} )\;.
\ee
and a simple calculation shows that  
 \be       \sum_{i=1}^N    \nabla_i=    \nabla_{\bf R}\;.
\ee
as is well known from classical mechanics.      (\ref{Ehr2})  then  simplifies  also:  
\be     \frac{d}{dt}  \brc \Psi   |  {\bf P}  |\Psi \ckt  = \frac{1}{i \hbar}   \sum_i  \brc \Psi   |  [{\bf p_i}, H]  |\Psi \ckt =    
 -  \brc \Psi   |  \sum_i  \nabla_i   V     |\Psi \ckt  =  -  \brc \Psi   |  \nabla_{\bf R}    V     |\Psi \ckt  \;. \label{New2}
\ee
(\ref{New11}) and (\ref{New2})  taken together look almost like Newton's equation.  
As the left hand side of  equation (\ref{New2})    is
\be    \frac{d}{dt}   {\bf P}_{CM}(t) \;;    
\ee
one would have to  show
\be    \brc \Psi   |  \nabla_{\bf R}    V     |\Psi \ckt   =     \nabla_{ {\bf R}_{CM}} V_{class}( {\bf R}_{CM})\;,     \label{RHS} 
\ee
with an appropriately defined classical potential   $V_{class}( {\bf R}_{CM})$,   
 in order to derive Newton's equation for a  pointlike particle of mass $M$ at   ${ {\bf R}_{CM}}$.
But  such a relation cannot be taken for granted, actually, as will be seen below. 

A natural definition for  $V_{class}( {\bf R}_{CM})$  is   
\be   V_{class}( {\bf R}_{CM})  =     \brc \Psi  |  V (  {\hat  {\bf r}}_1,  {\hat  {\bf r}}_2, \ldots   {\hat  {\bf r}}_{N-1},   {\bf R})   |  \Psi \ckt\;.   \label{CMV} 
\ee
Now  the wave function for the center of mass of a macroscopic body can be taken as an arbitrarily  narrow wave packet (Sec.~\ref{basic}).   The width of the  CM wave function packet  indeed corresponds to an  experimental measurement (preparation) of the  CM position
made with an ``arbitrary" precision, see   (\ref{Definite}),   with  $\Delta X_{CM}$  certainly much smaller than  body's size, $L_0$:
\be   \Delta  X_{CM}  \ll   L_0\;.    \label{evenif}
\ee

Actually,   the finite size of the body  $L_0$  affects    significantly  the  calculation of the  right hand side of the second Ehrenfest theorem,  (\ref{New2}), 
except in few special cases  which will be discussed    below,  in Sec.~\ref{special1}
and Sec.~\ref{special2}. In more  general cases, such as  those discussed in Sec.~\ref{general1},     the corrections  from the  Newton equation for a pointlike mass $M$,   (\ref{RHS}),    
arise in perturbation theory applied to Ehrenfest's equations.    These deviations  reproduce  the effects well known in classical physics for macroscopic bodies,  such as the gravitational tidal forces.
The related effects, for microscopic systems,   are well known and  discussed in many QM textbooks (e.g., deriving the polarization effects  and van der Waals forces  for atoms and molecules).    
   It is  nonetheless satisfactory  
  that our efforts to identify Newton's equations for a macroscopic body  in QM    (Sec.~\ref{Heisenberg} - Sec.~\ref{Decoherence}  and Sec.~\ref{Newton})      naturally leads to appropriate corrections due to the finite-size of macroscopic bodies, which are also known in classical physics.

\subsubsection{Justifying the use of  Ehrenfest's theorem  for macroscopic bodies  \label{subtle} }

Before embarking on the actual application of the Ehrenfest theorem for  macroscopic bodies in some concrete cases, we make a brief pause to discuss an apparently minor, but actually a fundamental,  question.  

One of the essential ingredients for the emergence of the classical physics is decoherence  \cite{Joos1}-\cite{Arndt1},  discussed in Introduction and 
in Sec.~\ref{Decoherence} in our context.    It means that the macroscopic body under consideration has $T>0$ and is  radiating: it is a mixed state, which looks like  (\ref{macrobody}), with many infrared photons emanating from it.   How can a  na\"ive treatment such as  in Sec.~\ref{EhrMac}   
in which the state of the macroscopic body is represented by  a pure-state wave function of $N$  atoms and molecules,  justified?   

There are at least two good  reasons why such an approach is indeed appropriate.  
The first  is the fact that  the angular distribution of the emitted photons can be considered random statistically,   and therefore regarded as   spherically symmetric. The average recoil from the  different emitted photons on the   CM  motion of the macroscopic body will cancel out, to become  negligible.  

     But perhaps more significantly,   
effects of the recoil are  negligibly small,   for an emission from excited atoms and molecules {\it  bound  in  a solid  (crystal)},     
due to the fact that   the recoil kinematics involves whole body's  large mass, $M$.  This phenomenon is well known: it  is the essence of the M\"ossbauer effect.  

Because of  (at least) these two eproperties of the infrared emissions from a macroscopic body,  the actual motion of the CM of a macroscopic body is not significantly affected by them.   The CM motion of the body described by each term of (\ref{macrobody})  is identical, thus we are entitled to study  
the CM motion of a macroscopic body,  treating it as if it were a pure state described by a wave function. 

  A moment of thought shows that this apparently minor and natural observation is,  actually, at the heart of the whole discussion.  
 The central question regards the paradoxical role played by certain microscopic quantum processes involved.  We have in mind the scattering  of the body  with air molecules, cosmic rays, and   other background fluxes,  in the case of ``environment-induced decoherence" processes  \cite{Joos1}-\cite{Arndt1},  and  the spontaneous  emission of infrared photons from a macroscopic body at finite temperature  in  our case,  (\ref{macrobody}),  or in  the  $C_{70}$  experiments \cite{C70}.    They are crucial  and responsible for decoherence, hence for the possibility that the CM of a macroscopic body possesses a unique classical trajectory. Nevertheless, for the purpose of finding out how it moves under,  e.g.,  gravitational forces or external electromagnetic forces, and of proving  that  it  indeed obeys Newton's equations, these microscopic physical processes represent inessential backgrounds and  (hopefully small) corrections. They must be neglected 
 in the first approximation for the derivation of Newton's equations from quantum mechanics.

\subsection{Gravitational potential on the  earth  \label{special1} }

As the first concrete example  let us  consider the case of external gravitational force on a massive body on the earth. The total potential is 
\be     V ({\bf r}_1,  {\bf r}_2, \ldots  {\bf r}_N) =   V_{int}({\hat  {\bf r}}_1,  {\hat  {\bf r}}_2, \ldots   {\hat  {\bf r}}_{N-1}) + V_{grav}( {\bf R})\;\label{formbis} 
\ee
where the first term is independent of  the CM position ${\bf R}$.   The Hamiltonian is the sum
\be  H= H_{CM} +   H_{int}\;,   
\label{TotHamBis}  \ee 
\be       H_{CM}=  \frac{{\bf P}^2}{2M} +    V_{grav}( {\bf R})\;;\qquad     H_{int}= \sum_i    \frac{{\hat  p}_i^2}{2\mu_i} +   V_{int}({\hat  {\bf r}}_1,  {\hat  {\bf r}}_2, \ldots   {\hat  {\bf r}}_{N-1}) \;,
\label{HamiltoniansBis}  \ee
where 
\be     V_{grav}( {\bf R})=g    \sum_i   m_i z_i  =      M  g   Z\;,  \label{gravPot}
\ee
where $g\simeq 980 \, {\rm cm}/{\rm sec}^2$ is the acceleration  on the surface of the earth.

The  form of the Hamiltonian  (\ref{formbis})   allows us  to take the wave function in the factorized form,  
\be      \Psi =    \Psi_{CM}({\bf R}) \psi({\hat  {\bf r}}_1,  {\hat  {\bf r}}_2, \ldots   {\hat  {\bf r}}_{N-1}) \;. 
\ee
The time independent Schr\"odinger equation 
\be   H \Psi = (  H_{tot}  +   H_{int} ) \Psi   =  E_{tot} \,  \Psi   \;, 
\ee
gives 
\be     H_{int}\,  \psi({\hat  {\bf r}}_1,  {\hat  {\bf r}}_2, \ldots   {\hat  {\bf r}}_{N-1})  =  E_{int}  \,\psi({\hat  {\bf r}}_1,  {\hat  {\bf r}}_2, \ldots   {\hat  {\bf r}}_{N-1}) \;;
\ee
\be  H_{CM}  \Psi_{CM}({\bf R})  =  E_{CM}  \Psi_{CM}({\bf R}) \;,
\ee
and 
\be     E_{tot}=    E_{CM}   +   E_{int}\;,  
\ee  
where   $H_{int}$,   $\psi({\hat  {\bf r}}_1,  {\hat  {\bf r}}_2, \ldots   {\hat  {\bf r}}_{N-1})$ and  $ E_{int}$ refer to the internal excitation modes
of the body under consideration
\footnote{In the case of certain  rigid macroscopic bodies (e.g., crystals),     $H_{int}$ and $E_{int}$  contain the  total angular momentum contribution of the rotational modes of the massive body as a whole (and the related energy), which  should  not be considered as the internal excitation modes.    In  the present work, however,  we will not separate these three degrees of freedom  from the internal excitations.} .

 (\ref{New2})  can be replaced  by
\be     \frac{d}{dt}  \brc \Psi_{CM}  |  {\bf P}  |\Psi_{CM} \ckt  =    -      \brc \Psi_{CM}  |  \nabla_{\bf R}   V_{grav}( {\bf R})     |\Psi_{CM} \ckt  \;. \label{New222}
\ee
and (\ref{PosMom}) and (\ref{CMV})  by
\be      {\bf P}_{CM}(t)   \equiv     \brc \Psi_{CM} |  {\bf P}  |\Psi_{CM}  \ckt\;;  \qquad    {\bf R}_{CM}(t)  \equiv     \brc \Psi_{CM}    |  {\bf R}  |\Psi_{CM}  \ckt \;,     \label{CMR2}
\ee
and 
\be   V_{class}( {\bf R}_{CM})  =   \brc \Psi_{CM}  | V_{grav} ({\bf R})   |  \Psi_{CM} \ckt\;, \label{CMV2} 
\ee
where   $\Psi_{CM}$  is an arbitrarily narrow  (see  (\ref{Definite})) wave packet describing the CM wave function,  centered at
${\bf R}_{CM}(t)$ and  ${\bf P}_{CM}(t)$.

For such a macroscopic body to obey Newton's law,    
\be      \frac{d}{dt}   {\bf P}_{CM}(t)   =   
 -        \nabla_{{\bf R}_{CM}}   V_{class}( {\bf R}_{CM}) 
 \;.  \label{New22222}  
\ee
we need an approximate equality, 
\be  \brc \Psi   |  \nabla_{\bf R}   V_{grav}({\bf R})     |\Psi \ckt \simeq         \nabla_{{\bf R}_{CM}}   V_{class}( {\bf R}_{CM})  \brc \Psi   |  \Psi \ckt     =     \nabla_{{\bf R}_{CM}}   V_{class}( {\bf R}_{CM})  \;.   \label{derivation}
\ee
Actually, for the gravitational potential on the earth,   (\ref{gravPot}),  one has  
\be  -   \nabla_{\bf R}   V_{grav}({\bf R})    =     -   \nabla_{\bf R}    M  g  Z =    -   M   g \,   {\bf  \hat   k} \;,
\ee
a constant    (where   $ {\bf  \hat   k} $ is the unit vector in the   positive $z$  direction).  Also, 
\be    V_{class}( {\bf R}_{CM}) =  V_{grav}( {\bf R}_{CM})= M g  Z_{CM}\;,
\ee
where the dispersion of  $ {\bf R}_{CM}$  has been neglected.
The derivation of Newton's equation from QM,   (\ref{New22222}),     is therefore  exact.     
No details about the internal structures  of the body  ($H_{int} $  and  $\psi({\hat  {\bf r}}_1,  {\hat  {\bf r}}_2, \ldots   {\hat  {\bf r}}_{N-1})$) are needed to get this result.

\subsection{Motion in a gravitational potential due to a distant mass \label{general1} }

Let us now consider the gravitational force  on a body with mass $M= \sum_i  m_i$, exerted by  a distant mass $M_0$ at ${\mathbf R}_0$.   
The gravitation potential is   in this case
\be   V_{grav} ({\bf r}_1,  {\bf r}_2, \ldots  {\bf r}_N) =    \sum_i    \frac{ G_N    \, m_i   M_0}{|{\bf r}_i  -   {\bf R}_0|}\;, 
\ee
where $G_N$ is Newton's  gravitational constant,
\be  G_N \simeq  6.674   \times  10^{-11}   \,  {\rm m}^3  {\rm Kg}^{-1}   {\rm s}^{-2}\;. 
\ee 
We assume that the gravitational force is weak as compared to the  internal  forces which bind the system into the macroscopic body. In other words,   in the total Hamiltonian,
\be    H =    H_{CM}   +   H_{int}    +    V_{grav} ({\bf r}_1,  {\bf r}_2, \ldots  {\bf r}_N) \;,     
\ee
\be   H_{CM} = \frac{{\bf P}^2}{2M}\;,\qquad    H_{int}= \sum_i    \frac{{\hat  p}_i^2}{2\mu_i} +   V_{int}({\hat  {\bf r}}_1,  {\hat  {\bf r}}_2, \ldots   {\hat  {\bf r}}_{N-1}) \;,\ee
the gravitational potential  $V_{grav}$  can be regarded as a perturbation.  $\mu_i$'s are the reduced masses (see (\ref{reducedmass}),(\ref{good})).

That means that   the   macroscopic body under consideration is a bound state described by  the time-independent Schr\"odinger equation,  
\be     H_{int}\,  \psi^{(0)}({\hat  {\bf r}}_1,  {\hat  {\bf r}}_2, \ldots   {\hat  {\bf r}}_{N-1})  =  E_{int}  \,\psi^{(0)}({\hat  {\bf r}}_1,  {\hat  {\bf r}}_2, \ldots   {\hat  {\bf r}}_{N-1}) \;.   \label{unpert1}
\ee
{\it Before}   taking into account the gravitational potential,  
 the wave function thus has a factorized form,  
\be      \Psi =    \Psi_{CM}({\bf R}) \psi^{(0)}({\hat  {\bf r}}_1,  {\hat  {\bf r}}_2, \ldots   {\hat  {\bf r}}_{N-1}) \;. \label{unpert2}
\ee
In (\ref{unpert1}) and (\ref{unpert2}) a suffix $(0)$    (``unperturbed")   is used to indicate that the wave function   $ \psi^{(0)}$  stands for the macroscopic body before
the gravitational potential is taken into account.

We assume  also  that  the distance from the body to the mass $M_0$  is much larger than the  linear  size of the body (the spatial extension of  $ \psi^{(0)}$), $L_0$.   We may then assume that  
\be      \forall i    \;,        \quad        | {\hat    {\mathbf r}}_i  |  =     |{\mathbf r}_i -  {\mathbf R}| \sim    L_0 \;,    \qquad    \frac{L_0 }{| {\mathbf R} -   {\mathbf R}_0 |}       \ll   1   \;.  \label{approx} 
\ee
Actually, some care is needed  in such an approximation.   The reason is that   ${\mathbf r}_i$'s and   $ {\mathbf R}$  in  (\ref{approx})  are position {\it  operators},  and as such, have no \`a priori definite values.  However,  we remember that  these relations   are to be   used between the bra and ket  of the  state  vector of the  macoscopic body, $\Psi$.  As  for  $ {\hat  {\bf r}}_i$'s,       their extensions are effectively limited by the support of  the  
 internal wave  function $ \psi^{(0)}({\hat  {\bf r}}_1,  {\hat  {\bf r}}_2, \ldots   {\hat  {\bf r}}_{N-1}),$   hence by the size $L_0$ of the body.   In the case of   the center of mass position  ${\mathbf R}$  it is a bit  subtler,  as $  \Psi_{CM}({\bf R}) $     satisfies a free Schr\"odinger equation.  The consideration  in Sec.~\ref{Heisenberg} and Sec.~\ref{Diffusion},  however allows us to use for the CM wave function $  \Psi_{CM}({\bf R}) $ a well-defined, narrow wave packet, with an extension much smaller than  $L_0$  (see (\ref{evenif})).  
The  approximation (\ref{approx})  is therefore justified.

The gravitational potential can now be multipole expanded in powers of   $1/ {|{\mathbf R}  -   {\bf R}_0|}$:
\bea    && V_{grav}  ({\bf r}_1,  {\bf r}_2, \ldots  {\bf r}_N) =    \sum_i    \frac{ G_N    \, m_i   M_0}{| {\hat   {\bf r}}_i  +   {\mathbf R}   -   {\bf R}_0|}
\nonumber  \\
&&   =       \frac{ G_N  M   M_0}{|{\mathbf R}  -   {\bf R}_0|}    -    G_N  \sum_i  \frac{  m_i   {\hat   {\bf r}}_i^2}{ |  {\mathbf R}   -   {\mathbf R}_0|^3} 
+   3 G_N \sum_i   \frac{  m_i   \left(  \hat  {\mathbf r}_i  \cdot ({\mathbf R} -  {\mathbf R}_0 )  \right)^2  }   { | {\mathbf R}   -   {\mathbf R}_0 |^5 }  
+ \ldots     \label{quadr}  
\eea
The first term corresponds to the gravitational potential for the pointlike mass $M$ at ${\mathbf R}$.

Note that the first, order $O((\frac{1}{|{\mathbf R}  -   {\bf R}_0|})^2 )$   correction term,   the ``dipole term",  
\be  -   \sum_i   \frac{ G_N \, m_i  {\hat  {\mathbf r}_i}  \cdot ( {\mathbf R} -  {\mathbf R}_0)     } { |  {\mathbf R}   -   {\bf R}_0|^3} 
\ee
is actually  absent in (\ref{quadr}),   as is well known for the gravity,   as    $\sum_i   m_i  {\hat  {\mathbf r}_i} =0$.

 The second and third term of  (\ref{quadr})   represent   the quadrupole  corrections,  the magnitude of which depends on the  internal  wave function, $\psi({\hat  {\bf r}}_1,  {\hat  {\bf r}}_2, \ldots   {\hat  {\bf r}}_{N-1})$, i.e., the non-spherically-symmetric form of the body.

An important point  to note however   is that  the quadrupole (and higher) corrections will appear, even if the unperturbed wave function $\psi^{(0)}$
(the body without the influence of the external gravity  due to the distant mass), is perfectly spherically symmetric.  In such a  case,  the contribution of the  second and the third term of   
(\ref{quadr})   on the right hand side of the second  Ehrenfest's theorem,  
\be      \frac{d {\bf P}_{CM} }{dt}=         
 -  \brc  \Psi   |  \sum_i  \nabla_i   V     |\Psi \ckt   
\ee
might  appear  to   vanish,    if the factorized wave function (\ref{unpert2}) is used, with a  spherically-symmetric  internal wave function  $ \psi^{(0)}({\hat  {\bf r}}_1,  {\hat  {\bf r}}_2, \ldots   {\hat  {\bf r}}_{N-1})  $.

Actually,   a nonvanishing quardupole  appears as the result of the   
perturbation  to the wave function: the deformation of the body due to the external gravitational force.   In order to see these effects,   it is necessary to study the consequence of the perturbation due to external gravity systematically in the context of the Ehrenfest theorem.  In Appendix~\ref{PertEh}   a standard perturbation theory  is  applied to Ehrenfest's theorem,   
to  derive  modifications to the force, order by order  in perturbation.

In order to apply perturbation theory to Ehrenfest's theorem,  as sketched in   Appendix~\ref{PertEh},  it is  however  
more convenient to split the Hamiltonian  differently.  Namely, we write  
\be       V_{grav}  ({\bf r}_1,  {\bf r}_2, \ldots  {\bf r}_N)=    \frac{ G_N  M   M_0}{|{\mathbf R}  -   {\bf R}_0|}   +   V_{mp}({\bf r}_1,  {\bf r}_2, \ldots  {\bf r}_N)\;,  \label{multipole1}
\ee
and consider the multipole expansion terms
\be    V_{mp}({\bf r}_1,  {\bf r}_2, \ldots  {\bf r}_N)  =    -    G_N  \sum_i  \frac{  m_i   {\hat   {\bf r}}_i^2}{ |  {\mathbf R}   -   {\mathbf R}_0|^3} 
+   3 G_N  \sum_i   \frac{  m_i   \left(  \hat  {\mathbf r}_i  \cdot ({\mathbf R} -  {\mathbf R}_0 )  \right)^2  }   { | {\mathbf R}   -   {\mathbf R}_0 |^5 }  
+ \ldots     \label{multipole2}  
\ee
as perturbation.   The total Hamiltonian is written as 
\be    H =  H_0 +    V_{mp} ({\bf r}_1,  {\bf r}_2, \ldots  {\bf r}_N) \;,\qquad H_0=  {\hat  H}_{CM}    +   H_{int} \;, 
\ee
\be   {\hat  H}_{CM}  = \frac{{\bf P}^2}{2M}  +     \frac{ G_N  M   M_0}{|{\mathbf R}  -   {\bf R}_0|}     \;,\qquad    H_{int}= \sum_i    \frac{{\hat  p}_i^2}{2\mu_i} +   V_{int}({\hat  {\bf r}}_1,  {\hat  {\bf r}}_2, \ldots   {\hat  {\bf r}}_{N-1}) \;,\ee
and now  only the multipole expansion terms $ V_{mp} $  are treated as perturbation.

Unperturbed system  $H_0$    is still described by a factorized wave function,  
\be      \Psi =   {\hat   \Psi}_{CM}({\bf R}) \psi^{(0)}({\hat  {\bf r}}_1,  {\hat  {\bf r}}_2, \ldots   {\hat  {\bf r}}_{N-1}) \;. \label{unpert2Bis}
\ee
where  the CM wave function satisfies now the Schr\"odinger equation for  a pointlike mass $M$
\be       {\hat  H}_{CM}      {\hat   \Psi}_{CM}({\bf R})    =     E_{CM}  \,  {\hat   \Psi}_{CM}({\bf R})    \;   \label{NewtonCM}
\ee
  (which is Newton's law,   see below, (\ref{zeroth})),   
whereas  the internal wave function is given  by (see (\ref{unpert1})),
\be     H_{int}\,  \psi^{(0)}({\hat  {\bf r}}_1,  {\hat  {\bf r}}_2, \ldots   {\hat  {\bf r}}_{N-1})  =  E_{int}  \,\psi^{(0)}({\hat  {\bf r}}_1,  {\hat  {\bf r}}_2, \ldots   {\hat  {\bf r}}_{N-1}) \;,\label{unpert1111}
\ee
 which describes  the bound state without the effects of gravity by the distant mass.

The first Ehrenfest theorem takes the classic form,
\be     \frac{d}{dt}  \brc \Psi   |  {\bf R}  |\Psi \ckt  =  \brc \Psi  | \frac {\bf P}{M}  |\Psi \ckt  \;; \qquad  {\bf P}= \sum_i  {\bf p}_i \;,\label{New111}
\ee
that is, 
\be      \frac{d}{dt}   {\bf R}_{CM} =      \frac { {\bf P}_{CM}}  {M} \;, \qquad     {\bf R}_{CM} \equiv   \brc \Psi   |  {\bf R}  |\Psi \ckt  \;; \quad   {\bf P}_{CM} \equiv   \brc \Psi   |  {\bf P}  |\Psi \ckt  \;;  \label{classic} 
\ee
to all orders of perturbation.

The left hand side of the second Ehrenfest's theorem also contains
$  \frac{d}{dt}   {\bf P}_{CM} $ to all orders of perturbation, see  Appendix~\ref{PertEh}.  The  right hand side, which can thus  be interpreted as the force,   has an expansion in powers of  perturbation.

Let us assume,  for definiteness,    that, before taking into account    $V_{mp}$,  our macroscopic body is described by a spherically symmetric internal wave function, $\psi^{(0)}$.

We transcribe the formulas  (\ref{Ehr222}) $\sim$ (\ref{force02})  in  Appendix~\ref{PertEh}  to our three-dimensional macroscopic body with $N$-atoms (or molecules),  
\be   F \to{\bf F}\;,  \qquad     \frac{d}{dx} \to    \nabla_{{\bf R}} \;, \qquad   V_0 \to    \frac{ G  M   M_0}{|{\mathbf R}  -   {\bf R}_0|}    +    V_{int}({\hat  {\bf r}}_1,  {\hat  {\bf r}}_2, \ldots   {\hat  {\bf r}}_{N-1}) \;;      \label{context1}
\ee
\be       H^{(0)} \to     {\hat  H}_{CM}    +   H_{int}   \;,      \qquad   V^{\prime}  \to       V_{mp} ({\bf r}_1,  {\bf r}_2, \ldots  {\bf r}_N) \;,   \label{context2}
\ee
and
\be   \psi^{(0)} \to   \Psi =   {\hat   \Psi}_{CM}({\bf R}) \psi^{(0)}({\hat  {\bf r}}_1,  {\hat  {\bf r}}_2, \ldots   {\hat  {\bf r}}_{N-1}) \;.      \label{context3}
\ee
To the zeroth order in  $V_{mp}$,   (\ref{force00}) says that
\be   \frac{d}{dt}   {\bf P}_{CM}  =   {\bf F}^{(0)}   = -       \bra  {\hat   \Psi}_{CM}| \nabla   \frac{ G  M   M_0}{|{\mathbf R}  -   {\bf R}_0|}  |  {\hat   \Psi}_{CM}  \ckt  
\simeq    -   \nabla_{{\bf R}_{CM}}     \frac { G  M   M_0}  {  |{\mathbf R}_{CM}  -   {\bf R}_0|}   \;, \label{zeroth}    
\ee
where we have taken into account the fact that for a macroscopic body the CM  wave function can be taken as a wave packet, as narrow as we wish.   
The equation (\ref{zeroth}), together with (\ref{classic}),   is just the Newton equation for a pointlike mass of $M$ at  ${\mathbf R}_{CM}$.

A quick inspection of   (\ref{force01})  in our context   (\ref{context1})-(\ref{context3})  shows   that the  effect of    $V_{mp}$   on the motion of the macroscopic body vanishes 
to first order, 
$ {\bf F}^{(1)}   =0$,  
because of the angular momentum conservation (or by use of the Wigner-Eckart theorem).

The first nonvanishing effects of $V_{mp}$  on the motion of the body    
 arise at the second order,  $ {\bf F}^{(2)}$,       see  (\ref{force02}).    The operator  $  \frac{dV_0 }{dx} $  with the translation 
  (\ref{context1})   is  a zero-rank spherical tensor with respect to the internal degrees of freedom    ${\hat  {\bf r}}_i$.  On the other hand,    the  first-order correction to the wave function 
 $|\psi^{(1)} \ckt$,   (\ref{wfcorr1}),   contains states of angular momentum $2$ or higher   (see (\ref{multipole2})).     $|\psi^{(1)} \ckt$  would be exactly  $\ell=2$ state if only the quadrupole term (explicit in   (\ref{multipole2}))    were kept,  but in general   contains all  higher angular momenta  $\ell=2,4,6,\ldots$.   
 The second-order wave function correction    $|\psi^{(2)} \ckt$   can be seen from  (\ref{wfcorr2}) to contain angular momentum states    $\ell=0, 2,4,6,\ldots$.  

Physically, the nonvanishing  effect of quadrupole (and higher) moments  on the  motion of the CM can be  interpreted as due to the first order  deformation of the  body  (first order
correction to the internal wave function), and the resulting nonvanishing elements  in the second-order correction to the force,   (\ref{force02}),   due to the now non-spherically-symmetric  distribution of the atoms and molecules inside the body.

 \subsubsection{Moon and the ocean tides on the earth}  
 
\begin{figure}[h]
\begin{center}
\includegraphics[width=5in]{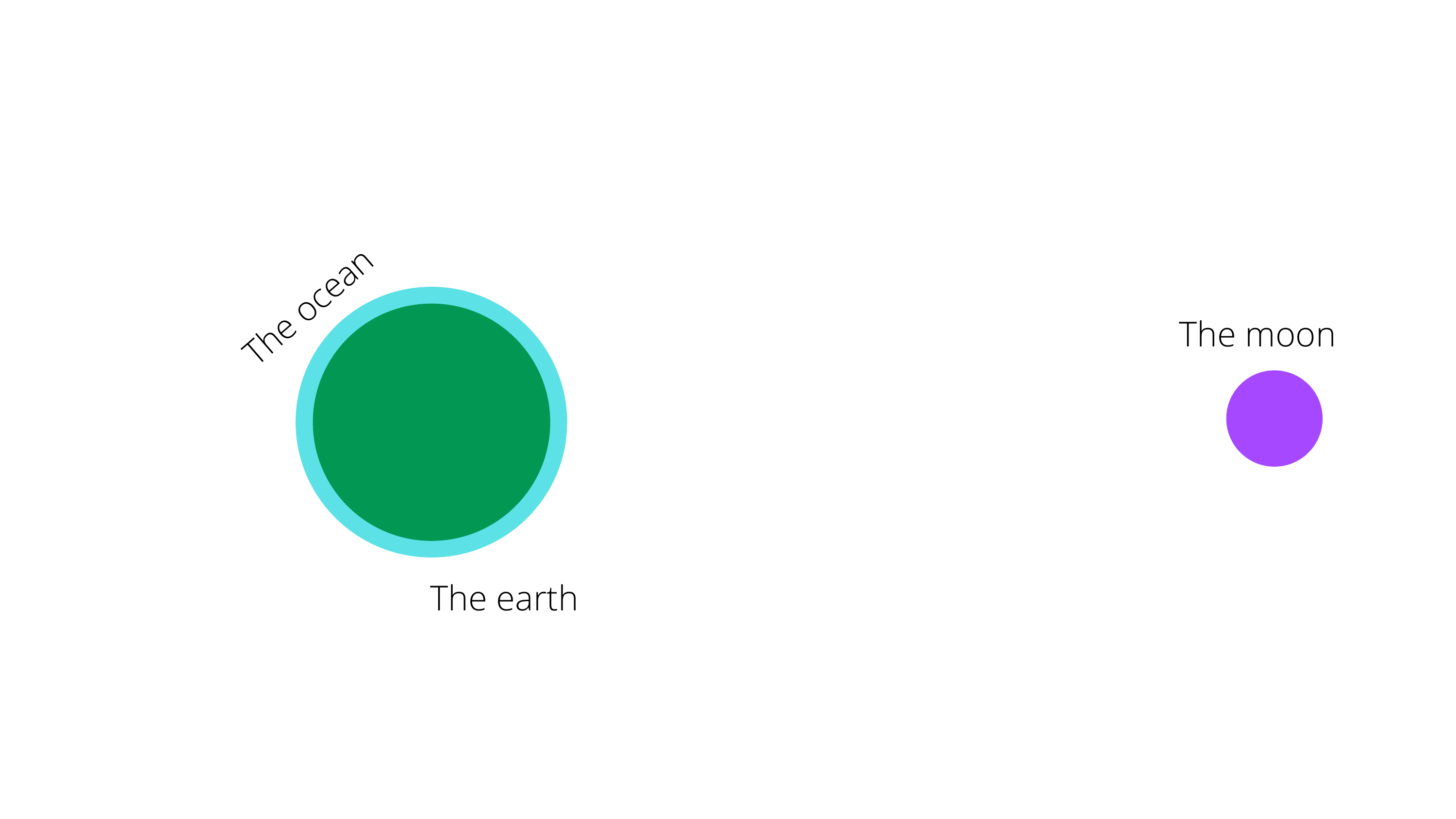}
\caption{\small  The earth and moon, in the approximation in which  the multipole expansion terms  $V_{mp}$ of  (\ref{multipole1}),  (\ref{multipole2})  are neglected.   The earth is assumed to be 
perfectly spherically symmetric, including its ocean.  }
\label{EarthMoon1}
\end{center}
\end{figure}

\begin{figure}[h]
\begin{center}
\includegraphics[width=5in]{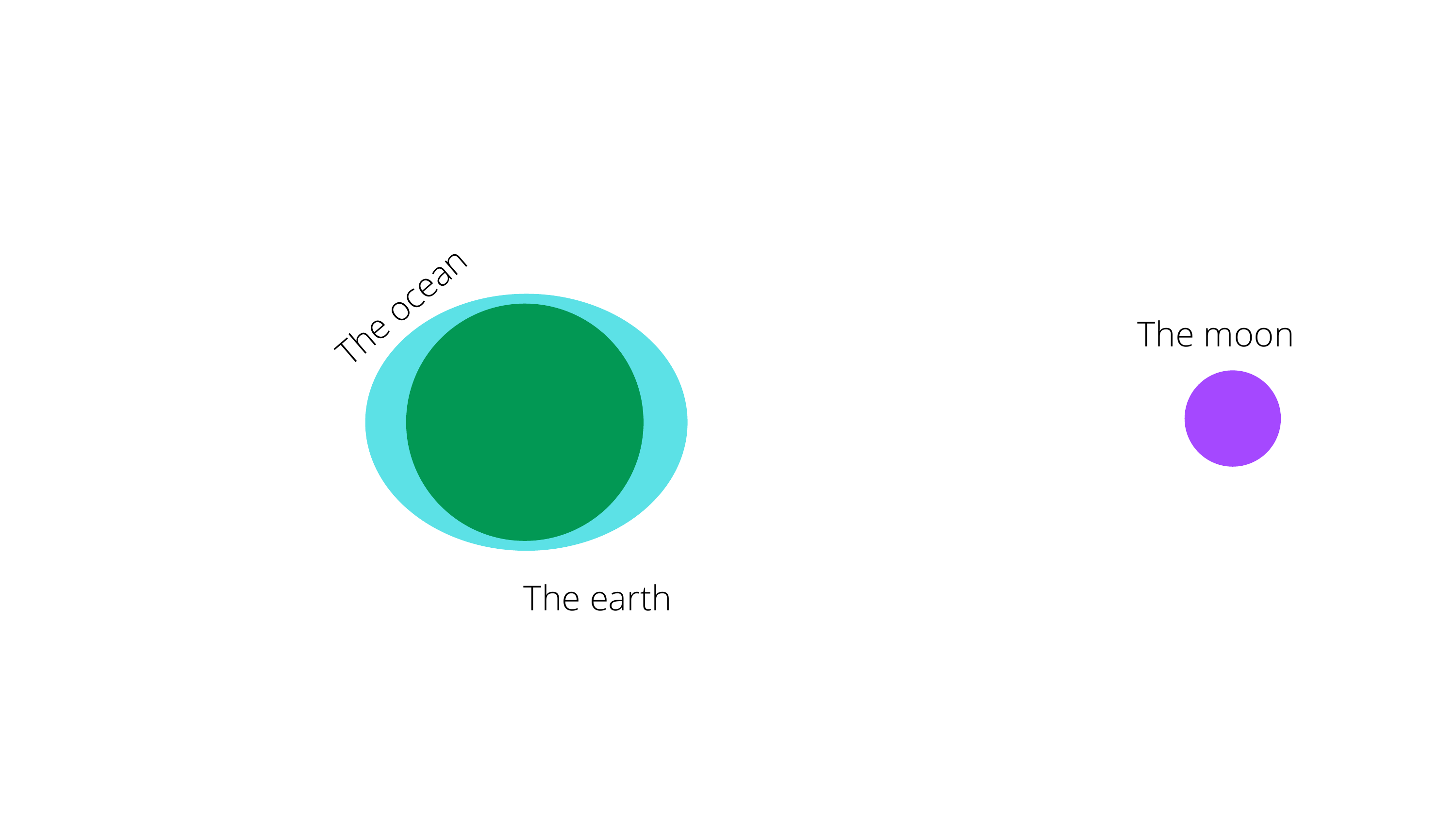}
\caption{\small  The earth's wave function is now deformed by the perturbation, $V_{mp}$, (\ref{multipole2}).  The main effect is in the change of distribution of the water molecules which are more loosely bound than the atoms and molecules in the earth's crust.  }
\label{EarthMoon2}
\end{center}
\end{figure}

 A familiar phenomenon related to these considerations  is the ocean tides  on  the earth, due to the gravitational force of the moon. 
 A caricature situation of  the earth and moon, to the zeroth-order  in  $V_{mp}$,  is shown  in Fig.~\ref{EarthMoon1}.  The earth, including the ocean, is assumed to have  a spherically symmetric form, before  $V_{mp}$ is taken into account. 
 The earth and moon  move according to Newton's equations between the pointlike masses  $M_{earth}$ and $M_{moon}$,  see (\ref{zeroth}), (\ref{classic}).
 When the multipole perturbation $V_{mp}$ is taken into account,  the  earth gets deformed.   Note that the deformation of the  earth's  wave function  is mainly  due to 
 the shifts of the water molecules  in the ocean:    their bindings are  much weaker than the binding of the molecules and atoms in the solid earth crust.  The earth 
 and moon now  look like  Fig.~\ref{EarthMoon2}.  This gives rise to the daily high and low ocean tides  (roughly twice each)  on the earth \footnote{According to Feynman \cite{Feynman} it was Newton who has first correctly identified  the cause of the daily ocean tides. }. 
 
 Though done with a crude,  simplifying assumptions,   this discussion may be regarded   as the first  derivation of the  familiar bidiurnal
ocean tides,   by a direct use of  the Schr\"odinger equation  for a  macroscopic body (the earth)   described by  a wave function with  $N\sim 10^{51}$  atoms !
 
This is, of course, only a statement of principle.  Fluid dynamics \cite{Tritton,Landau6}  is based on the idea that each element of the fluid, whose volume is much larger than the size of the molecules composing it, can be regarded as a classical particle. Thus a quantitative analysis of the ocean tide may be done classically \cite{Stewart}.

\subsection{Harmonic oscillator potential     \label{special2} }

Next consider the case of a hypothetical harmonic-oscillator potential acting  uniformly on all the components of the body. The potential is given by 
\be     V = \sum_{i=1}^N       \frac{   m_i \omega^2     {\mathbf r}_i^2 }{2}   +   V_{int}({\hat  {\bf r}}_1,  {\hat  {\bf r}}_2, \ldots   {\hat  {\bf r}}_{N-1}) \;,  
\ee
where $ V_{int}$ describes the interactions responsible for binding the body, as before. 
In this case the Hamiltonian does not simply factorize as a sum of the CM  part and the internal Hamiltonian,  as  in   (\ref{HamiltoniansBis}). Nevertheless, Ehrenfest's theorem simplifies again. 

The first Ehrenfest theorem reads,
\be     \frac{d}{dt}  \brc \Psi   |  {\bf R}  |\Psi \ckt  =  \brc \Psi  | \frac {\bf P}{M}  |\Psi \ckt  \;; \qquad  {\bf P}= \sum_i  {\bf p}_i \;,\label{New11111}
\ee
that is, 
\be      \frac{d}{dt}   {\bf R}_{CM} =      \frac { {\bf P}_{CM}}  {M} \;.
\ee
The second simplifies as well: 
\bea   &&        \frac{d {\bf P}_{CM} }{dt}=        \frac{d}{dt}  \brc \Psi   |  {\bf P}  |\Psi \ckt  =   \frac{1}{i \hbar}   \sum_i  \brc \Psi   |  [{\bf p_i}, H]  |\Psi \ckt =    
 -  \brc  \Psi   |  \left( \sum_i  \nabla_i   V  \right)    |\Psi \ckt   = \nonumber \\
 &&    -   \brc \Psi   |   \sum_{i=1}^N     m_i \omega^2     {\mathbf r}_i    |\Psi \ckt   =   -  M  \omega^2    \brc \Psi | {\mathbf R}      |\Psi \ckt =     -  M  \omega^2   {\mathbf R}_{CM}    \;. \label{New222HO}
\eea 
Note that $   \sum_{i=1}^N    \nabla_i=    \nabla_{\bf R}$ does not act on the  internal  interaction potential,     $ V_{int}$. 
The center of mass of the whole body  obeys Newton's equation for a classical harmonic oscillator of frequency $\omega$.

Again,  the knowledge about the internal forces which bind  the body
is not required.

\subsection{External electromagnetic  fields    \label{EM}   }

It is in principle straightforward to generalize these discussions to the case of a macroscopic body  in an external electromagnetic fields,  especially 
in the limit the body can be considered pointlike,  and under the uniform and constant and  slowly varying  electromagnetic fields.

For  a macroscopic particle of total charge $Q$
 the Hamiltonian is given by   
\be    H \to  H=      \sum_i   \left[  \frac{({\bf p}_i - \frac{e_i}{c} {\bf A}({\bf r}_i) )^2}{2m_i}  +  e_i \Phi({\bf r}_i) - {\bf \mu}_i \cdot {\bf B}({\bf r}_i)    \right]   +    V_{int}({\hat  {\bf r}}_1,  {\hat  {\bf r}}_2, \ldots   {\hat  {\bf r}}_{N-1}) \;, \label{macroEM}
\ee
\be  \sum_i e_i = Q\;; \qquad     {\bf B} =   \nabla \times   {\bf A};\qquad      {\bf E}   =   \nabla   \Phi({\bf r})\;,   
\ee
and $\mu_i$ is the intrinsic magnetic moment of the $i$-th constituent. 
The interactions with the {\it  external} electromagnetic fields are  included in the expression inside the square brackets;    the last term of (\ref{macroEM})  
represents all other interactions among the atoms and molecules describing the bound state.
 $V_{int}$ is assumed to have no dependence on the CM coordinate,
 \be \nabla_{\bf R} V_{int} =0\;: \label{noR}
 \ee
  the deviation  from the free motion for the  CM  of the body under study, is 
due only to the external electromagnetic fields. 
 For simplicity, we have left  the dependence of   $V_{int}$ on  the spin variables and on the momenta  ${\hat  {\bf p}_i}$   (such as in the relativistic corrections and spin-orbit interactions)   implicit.

\subsubsection{Motion in  a constant electric field \label{constE}}

Let us start with the simplest case of a constant electric field,  ${\bf E}$:  we take 
\be      \Phi({\bf r}) =   {\bf E} \cdot {\bf r}\;, \qquad   {\bf A}=  {\bf B}=0\;. 
\ee 
The first Ehrenfest equation for the CM position reads simply
\bea     \frac{d}{dt}  \brc \Psi   |  {\bf R}  |\Psi \ckt  &=&  \sum_i  \frac{d}{dt}  \brc \Psi   |  \frac{ m_i {\bf r}_i}{M}    |\Psi \ckt  =      \frac{1} {i\hbar}  \sum_i     \brc \Psi   |    \frac{ m_i}{M}      [{\bf r}_i, H]  |\Psi \ckt  \nonumber \\
&=&         \brc \Psi   |    \frac{1}{M}  \sum_i     {\bf p}_i   |\Psi \ckt     =    \frac{1}{M}   \brc \Psi   |   {\bf P} | \Psi \ckt    \;.   \label{asin}    
\eea   
The second equation is 
\be     \frac{d}{dt}  \brc \Psi   |  {\bf P}  |\Psi \ckt  = \frac{1}{i \hbar}   \sum_i  \brc \Psi   |  [{\bf p_i}, H]  |\Psi \ckt     \;, 
\ee
but
\be   \frac{1}{i \hbar}   \sum_i  \brc \Psi   |  [{\bf p_i},  \sum_j  e_j \Phi({\bf r}_j)  ]    |\Psi \ckt   = -   \sum_i    \brc \Psi |     e_i  {\bf E}    |\Psi \ckt    = - \, Q\,  {\bf E}\;,  
\ee
where $Q= \sum_i e_i$ is the total charge of the body, whereas  
\bea        \frac{1}{i \hbar}   \sum_i  \brc \Psi   |  [{\bf p_i},  V_{int}]   |\Psi \ckt   &=&   -   \sum_i  \brc \Psi   |  [{\nabla_i},  V_{int}]   |\Psi \ckt  = -  \brc \Psi   |  [{\nabla_{\bf R}   },  V_{int}]   |\Psi \ckt   \nonumber  \\
  & = &  
      -  \brc \Psi   |  (\nabla_{\bf R} V_{int} ) |\Psi \ckt  =0\;
\eea
(see (\ref{noR})). 
Therefore  a narrow wave packet for the CM, centered at ${\bf R}_{CM}, {\bf P}_{CM}$,    obeys 
\be          \frac{d}{dt}  {\bf R}_{CM} =      \frac{1}{M}   {\bf P}_{CM}\;, \qquad     \frac{d}{dt}  {\bf P}_{CM} =  - \, Q\, {\bf E}\;,
\ee
as expected.   If the electric field is due to a charge $Q_0$ at  ${\bf R}_0$ at a distance  from ${\bf R}_{CM}$  much greater than the body's size,  $L_0$,  one finds Coulomb's law, 
\be     \frac{d}{dt}  {\bf P}_{CM} =    {\bf F} =     -   \nabla_{{\bf R}_{CM}} \frac{ Q  Q_0}{|{\bf R}_{CM} -  {\bf R}_0|}\;
\ee
in the first approximation  (in which the body can be regarded as  pointlike),   by repeating an argument similar to the one which led to  (\ref{zeroth}).

 \subsubsection{Motion in a constant magnetic field}

Let us next  consider an external, uniform constant magneti field ${\bf B} $.   The vector potential can be taken as 
\be     {\bf A}({\bf r})    =   \frac{1}{2}   \, {\bf B} \times  {\bf r}  \;.     \label{vectorpot}
\ee  
The first Ehrenfest theorem  now  reads 
\be     \frac{d}{dt}  \brc \Psi   |  {\bf R}  |\Psi \ckt  =  
  \brc \Psi   |    \frac{1}{M}  \sum_i   (  {\bf p}_i  -    \frac{e_i}{c}   {\bf A}({\bf r}_i) )   |\Psi \ckt   =     
  \brc \Psi   |    \frac{1}{M}  \sum_i   \left(  {\bf p}_i  -    \frac{e_i}{2 c}  ( {\bf B} \times  {\bf r}_i  )   \right)   |\Psi \ckt   
\ee
The expectation value of the terms containing  $ {\bf r}_i $  vanish due to parity,
and one gets  
\be
   \frac{d}{dt}  \brc \Psi   |  {\bf R}  |\Psi \ckt  =  \frac{1}{M}   \brc \Psi   |   {\bf P} | \Psi \ckt \;;\qquad     \frac{d}{dt}   {\bf R}_{CM}  =     \frac{1}{M}  {\bf P}_{CM} \;.
\ee
The second Ehrenfest equation  is 
\bea     \frac{d}{dt}  \brc \Psi   |  {\bf P}  |\Psi \ckt  &=&    \frac{1}{i \hbar}   \sum_i  \brc \Psi   |  [{\bf p_i}, H]  |\Psi \ckt     \nonumber \\
 &=&  
 \frac{1}{i \hbar}   \sum_i  \brc \Psi   |  [{\bf p_i},              \sum_j    \frac{({\bf p}_j - \frac{e_j}{c} {\bf A}({\bf r}_j) )^2}{2m_j}  ]              |\Psi \ckt   \;.\eea
By using (\ref{vectorpot}),  one gets after some calculation,  
\be   \frac{1}{i \hbar}   \sum_i  [{\bf p_i},              \sum_j    \frac{({\bf p}_j - \frac{e_j}{c} {\bf A}({\bf r}_j) )^2}{2m_j}  ]  =     \sum_i   \frac{e_i}{ m_i c}  ({\bf p}_i -   \frac{e_i}{c} {\bf A}({\bf r}_i) ) \times  {\bf B}\;,
\ee
where a vector notation is used throughout,   in order to avoid confusion between the space indices and  the particle indices, $i, j$.    We now recognize 
\be   {\bf  j}_i =  \frac{e_i}{ m_i}  ({\bf p}_i -   \frac{e_i}{c} {\bf A}({\bf r}_i) )  
\ee 
as the familiar gauge-invariant (and conserved) electric current operator of the $i$-th constituent atom or molecule in an external magnetic field
   Writing now the total current as
\be    {\bf J}({\bf R}_{CM}) =     \brc \Psi   |  \sum_i {\bf  j}_i      |\Psi \ckt   \;,   
\ee  
one finds for the  CM  motion, 
\be         \frac{d}{dt}   {\bf P}_{CM}  =    \frac{1}{c}   {\bf J}   \times  {\bf B}\;,  \label{Lorentz}  
\ee
which is the Lorentz force.  
 
  \subsubsection{A macroscopic particle with a magnetic moment   \label{SGClassic}}

Let us now consider a  magnetic field slowly varying in space.  More precisely, we assume that the  space variation of ${\bf B}({\bf r})$  within the body under consideration, is negligible. 
We shall take 
\be       \nabla_{{\bf r}_i}  {\bf B}  =    0\;, \label{constB}
\ee 
as long as ${\bf r}_i$ is within the  support of the internal wave function  (which is a region of linear size $L_0$).
Note that under these circumstances the derivation of the Lorentz force  above continues to be valid:  the formula (\ref{Lorentz})   holds also for 
non constant magnetic field    ${\bf B} $.

To analyze this problem,  we first modify the kinetic terms in $H$  in (\ref{macroEM})  as
\be        \sum_i  \frac{({\bf p}_i - \frac{e_i}{c} {\bf A}({\bf r}_i) )^2}{2m_i}  \simeq \sum_i \left[    \frac{{\bf p}_i ^2}{2 m_i}   -  
\frac{e_i}{4  m_i  c }   \{  {\bf p}_i \cdot  ({\bf B} \times {\bf r}_i) +    ({\bf B} \times {\bf r}_i) \cdot  {\bf p}_i \}  \right]  \label{this}
\ee
where we dropped the terms proportional to the square of ${\bf A}$, and
  we have written  the vector potential 
   as 
\be     {\bf A}({\bf r}_i)    =   \frac{1}{2}   \, {\bf B}  \times  {\bf r}_i  \;,  \qquad  (i=1,2,.\ldots, N)\;,     \label{vectorpotBis}
\ee
even though now  ${\bf B}$ depends on the CM position ${\bf R}$.    To see how this step is justified,  consider  
\be   \nabla_{\bf r}  \times   \frac{1}{2}  ({\bf B}  \times {\bf r})  =     \frac{1}{2}   ({\bf r} \cdot \nabla)\, {\bf B}  + {\bf B}\;, \label{firstterm}
\ee
but the first term is negligible,  as long as this expression is  used inside  the expectation value,  
\be    \brc   \Psi |  \cdots   | \Psi \ckt \;
\ee
   (see (\ref{constB})).

   The right hand side of  (\ref{this}) can be seen to be equal to 
\be     \sum_i \left[    \frac{{\bf p}_i ^2}{2 m_i}   -  
\frac{e_i}{2  m_i  c }    {\bf B}  \cdot  ( {\bf r}_i  \times  {\bf p}_i)   \right] =   
  \sum_i \left[    \frac{{\bf p}_i ^2}{2 m_i}   -  
\frac{e_i}{2  m_i  c }  {\bf B}  \cdot  {\bf \ell}_i \right]  \;,    \label{}
\ee
where  ${\bf \ell}_i $  is the orbital angular momentum operator associated with the $i$-th constituent of the body.   We recognize 
\be    \frac{e_i    {\bf \ell}_i  }{2  m_i  c }  
\ee
as the magnetic moment produced by the orbital motion of the charged constituent (Bohr's magneton). By  denoting the total magnetic moment of the body as
\be          \sum_i   \langle \Psi |    ( \mu_i +     \frac{e_i    {\bf \ell}_i  }{2  m_i  c } ) |\Psi \rangle  =   \mu\;,  
\ee
we find  
\be      \frac{d}{dt}   {\bf P}_{CM}  =    -     {\nabla_{{\bf R}_{CM}} }  \,  ( \mu \cdot {\bf B} )\;,  \label{NoSG}
\ee
a familiar classical equation.

 \subsubsection{General consideration \label{general}}

One might think that the generalization of these results to include other effects, such as polarization,  charge conduction, etc., is quite a straightforward task.   
However, a moment of  reflection convinces us that  to try to reproduce the whole richness of electromagnetic phenomena  of condensed matter  (see for instance, \cite{Landau8}) and classical electromagnetism, 
starting from the Ehrenfest theorem for a macroscopic body with $N\sim 10^{26}$ atoms and molecules, is quite another matter.     The problem is that the reaction of the body to the external electromagnetic fields depends on the nature of the body, whether it is a conductor or a dielectic, or something else; in other words,  the full details of the internal Hamiltonian   and the internal  wave function $\psi({\hat  {\bf r}}_1,  {\hat  {\bf r}}_2, \ldots   {\hat  {\bf r}}_{N-1})$   describing the body, come into play.  In simple terms, what happens is that  each charged particle, electron, atomic nuclei, molecules, ions, react differently to the external electromagnetic field, and deform the internal structure of the body,  
according to the way and strength with which they are bound in the body. All of the detailed spin-dependent effects also will  come into play. 

Under such circumstance the analysis will quickly become very difficult;  the task requires    a systematic classification of different  physics 
cases, and an elaborated work organization and preparation. 

One has the feeling that what has been done in Sec.~\ref{constE} -  Sec.~\ref{SGClassic}  amounts to the  (though necessary) very first few steps towards climbing up Mt. Everest.

At first sight  the case of the external gravitational  forces considered in Sec.~\ref{special1} and Sec.~\ref{general1} might  appear exceptional.   
  Due to the universal attractive nature, proportional to the mass,  of the gravitational forces, most of the details of the internal interactions and wave functions drop  out of the equations.   Each constituent, small or large, inside the body reacts and accelerates in an identical manner under  the external gravitational force.  This explains the fact that  a macroscopic body, e.g., a metal piece,  which contain in its interior the electron gas which exhibits quantum-mechanical collective behavior even at room temperatures,  or even stars and planets with $N\sim  10^{60}$ atoms and molecules,  
 obey  Newton's law for its CM motion, as it is known empirically and as it has been shown in Sec.~\ref{special1}, Sec.~\ref{general1}.
 No doubt, this simplicity was the historical reason why classical mechanics was first discovered, mainly in the context of motions under gravitational forces. 
 
 Nevertheless, once  we go beyond the point-particle limit, the finite-size effects associated with the internal wave functions describing a large (typically, huge)  number $N$ of constituent atoms and molecules, come into play.  In other words, the entire richness of classical physics must in principle to be explained from quantum mechanics,  together with classical electromagnetism.    The general remark made above thus applies also here.

\subsection{Canonical equations for many-(macro-)-body systems  \label{canonical}}

So far, the discussion was limited to the case of a single macroscopic body under some external potential,  a gravitational, harmonic 
or electromagnetic potential.  
We sketch here  the generalization of these results to the case of  many-(macro-) body system made of $S$
macroscopic bodies  in the vacuum and at finite temperatures,  as described in Sec.~\ref{basic}.  We assume that each of the macroscopic bodies 
is described by an arbitrarily narrow wave packet both in the position and momentum for its CM, whereas  their sizes  $(L_0^{I})$ (the support of their internal wave function) are negligible as compared with the distances   $ |{\bf R}_{CM\,I} -   {\bf R}_{CM\,J}|$.

We assume that the  total  Hamiltonian  has the form, 
\be  H= H_{CM} +   \sum_I   H_{I\,int}\;,   
\label{BigHam}  \ee 
where the their centers of mass are described by
\be       H_{CM}=\sum_{I=1}^S  \frac{{\bf P}_I^2}{2M} +    V ({\bf R}_1, {\bf P}_1, {\bf R}_2, {\bf P}_2, \ldots, {\bf R}_S, {\bf P}_S)\;;
\qquad  {\bf P}_I= - i \hbar \nabla_{{\bf R}_I} 
\ee
and the internal Hamiltonians, 
\be        H_{I\,   int}= \sum_{i=1}^{N_I-1}    \frac{{\hat  p}_{I\,i}^2}{2\mu_{I,i}} +   V_{int}({\hat  {\bf r}}_{I,1},  {\hat  {\bf r}}_{I,2}, \ldots   {\hat  {\bf r}}_{I, N_I-1}) \;.
\label{BigInt}  \ee
describes the  binding  of the $I$-th macroscopic body.
The potential $V$  might contain  gravitational or Coulomb forces among the bodies  (as discussed in  Sec.~\ref{basic}), as well as 
higher-order effects  such as  the tidal effects, dipole-dipole interactions,  van-der-Waals forces,  etc.,  among the bodies.

Clearly  the Hamiltonian (\ref{BigHam}) is to be understood as a sort of effective Hamiltonian, 
whose derivation is left  for separate work;   here  it  is just  a (reasonable)  assumption we adopt.

Such a Hamiltonian would allow for the wave function of  factorized form,  
\be  | \Psi \ckt   =        | \Psi_{CM} \ckt     \prod_I    | \psi_I \ckt   
\ee
where
\be      H_{I\,   int}\,    \psi_I({\hat  {\bf r}}_{I,1},  {\hat  {\bf r}}_{I,2}, \ldots   {\hat  {\bf r}}_{I, N_I-1})  =  E_{I\, int}  \,\psi_I({\hat  {\bf r}}_{I,1},  {\hat  {\bf r}}_{I,2}, \ldots   {\hat  {\bf r}}_{I, N_I-1}) \;;
\ee
\be      i \hbar  \frac{d}{dt}     | \Psi_{CM} \ckt   =     H_{CM}    | \Psi_{CM} \ckt   =    E_{CM}   \,| \Psi_{CM} \ckt  \;.  
\ee

The Ehrenfest theorem for ${{\bf R}_I} $  reads 
\bea        \frac{d}{dt}    {\bf R}_{CM\,I} &=&   
  \frac{d}{dt}  \brc \Psi   |  {\bf R}_I   |\Psi \ckt  =      \frac{1} {i\hbar}     \brc \Psi_{CM}   |    [{\bf R}_I, H_{CM}]  |\Psi_{CM} \ckt =      \brc \Psi_{CM}     |    ( \nabla_{{\bf P}_I}  H_{CM} )  |\Psi_{CM}  \ckt \nonumber  \\
&=  &
\nabla_{{\bf P}_{CM\, I}}    H_{CM}({\bf R}_{CM\,1}, {\bf P}_{CM\,1}, {\bf R}_{CM\,2},  \ldots,  {\bf P}_{CM\,S})\,,  
 \label{canon1}
\eea
whereas the one for ${{\bf P}_I} $ gives 
\bea      \frac{d}{dt}    {\bf P}_{CM\,I} &=&      \frac{d}{dt}  \brc \Psi   |  {\bf P}_I   |\Psi \ckt  =      \frac{1} {i\hbar}    \brc \Psi_{CM}   |    [{\bf P}_I, H_{CM}]  |\Psi_{CM}   \ckt =    -    \brc \Psi_{CM}       |  ( \nabla_{{\bf R}_I}  H_{CM} )  |\Psi_{CM}  \ckt \nonumber  \\
&=   &
-   \nabla_{{\bf R}_{CM\, I}}    H_{CM}({\bf R}_{CM\,1}, {\bf P}_{CM\,1}, {\bf R}_{CM\,2},  \ldots,  {\bf P}_{CM\,S})\,,  
\label{canon2}
\eea
for $I=1,2,\ldots, S$.  In both  (\ref{canon1}) and  (\ref{canon2}), in the first and last steps  the center of mass position and momentum operators inside the expectation values    are  replaced by  the corresponding position and momentum centers of the (arbitrarily) narrow  wave packet of the $I$-th body
(Sec.~\ref{basic}).  Also,  the  expectation value of the Hamiltonian operator 
in such a narrow wave packet is approximated by a classical Hamiltonian.

 (\ref{canon1}) and  (\ref{canon2}) are  the familiar  canonical Hamilton equations of motion,    usually written in a compact form
 ($s$ being  the number of the degrees of freedom) 
 \be    \frac{d q_i}{dt} =  \frac{\de H}{\de p_i}\;;  \qquad    \frac{ d p_i}{dt} = -  \frac{\de H}{\de q_i}\;,    \qquad i=1,2,\ldots,  s\;.
 \ee

\section{Quantum versus classical worlds  \label{QCL}}

\subsection{Body-temperature      \label{temperature} }

Throughout the discussion of this work,     
there is clearly no sharp boundary between the quantum-mechanical and classical behavior of any given system as its temperature is varied.    A rough estimate of  the temperatures   $T\sim T_0$   under which the  microscopic, mesoscopic or macroscopic  system with a characteristic, lowest-excitation frequency 
$\nu$,    remains in its quantum-mechanical lowest  states,     is given by
\be    T  \lesssim      T_0\;,    \qquad  T_0  =     \frac{h \nu }{k_B}  =      \nu   \cdot  4.8   \cdot  10^{-11}   \, K  \;.   \label{temp} 
\ee  
The characteristic frequency  $\nu$ depends on many factors of the system, the size,  the mass, $N$, the atomic,  molecular or collective excitation energies, etc.

\begin{table}[h]
  \centering 
  \begin{tabular}{|c |c|c| c|c|c|}
\hline
   Systems      &               size      &     $  N  $   &    $\nu$ &  $T_0$      &            Ref.         \\   
 \hline 
  Macro pendulum (LIGO)      &     $10.8$ Kg     &  $\sim 10^{27} $    &    $100$     &   $ \sim 1.4  \cdot  10^{-6}  K$       &       \cite{LIGO}       \\
 \hline 
  Nanomech.  resonator     &      $10^{-13}  \, {\rm cm}^3$       &  $ \sim 10^{11}   $  &   $\sim  2\cdot 10^7$    &      $1  \, mK$       &       \cite{Aaron}    \\
 \hline
     Micromech. drum    &     $\sim  \mu $ m     &   $\sim 10^8 - 10^{10} $    &    $10^9$    &      $0.1 K$     &     \cite{LaHaye}            \\
 \hline
   $C_{70} $     &     $ 70$  {\rm  \AA} &  $70$     &     $10^{14}$    &       $\sim 3000  K$           &      \cite{C70}    \\
 \hline
\end{tabular}  
\caption{\footnotesize    A small sample of the experiments aimed at demonstrating the quantum mechanical behavior of a mesoscopic or macroscopic system at sufficiently low temperatures.   The size (the volume, linear dimension  or weight) and the number of atoms/molecules  cited are a crude estimate made from the lines of the articles by the present author:  any lack of precision is to be attributed to him, not to the authors of the works cited.
    }\label{classquant}
\end{table}

Some mesoscopic or macroscopic systems have been  studied recently for the purpose of revealing their quantum-mechanical behaviors
at low temperatures \cite{C70}, 
\cite{C60}-\cite{Brand}.
A small  sample of these experiments is summarized in  Table~\ref{classquant}, together with a few of the characteristics of the system considered and the relevant temperatures.
For  $N$ (the number of atoms/molecules)  spanning  over  many orders of magnitude,  the rule-of-thumb  estimate  (\ref{temp}) for the  typical temperature which devides 
quantum-mechanical and classical behaviors of the system,   seems to be  basically  valid.  

  In the case of  the LIGO  pendulum
of reduced mass $2.7$ Kg  \cite{LIGO},   the quoted  temperature of   $1.4  \,  \mu K$  refers to the  occupation numbers of about  $200$.  In the case of a nanomechanical resonator of \cite{LaHaye},  the quoted temperature   of $56\,mK$ refers to the   quantum excitation levels up to  $\sim 58$.    In a most 
 interesting  case of  the work by O'Connel et. al. \cite{Aaron},  the achieved temperature is  roughly consistent with the  system being  in its quantum ground state (no phonon excitations), as verified experimentally by coupling the system to a squid  qubit.
 
  For fullerene molecule  $C_{70}$,   the frequency $\nu$ is estimated from the known molecular  excitation energies, and  $T_0$  given by Arndt et. al.  
 \cite{C70}  is the temperature of the  $C_{70}$  beam  at which the interference effects disappear.  The temperature is raised by
 increasing the laser power bombarding  the $C_{70}$'s  before they enter the interferometer.   It is  remarkable  that by varying continuously the laser power 
 the authors  of \cite{C70}  have been  able to  interpolate  smoothly  between  the pure-quantum-state  $C_{70}$  and radiating, mixed-state ``classical" $C_{70}$.

The relatively higher temperatures needed in the case of  
$C_{70}$   as compared to the other macroscopic quantum-classical states in Table~\ref{classquant} reflects the fact that  here the excitations relevant are  the  $O(eV)$  {\it  molecular} excitations, whereas 
in other examples  the characteristic frequency (and the body  temperature)  refers to the collective, phonon excitation modes, or even to  macroscopic oscillation modes.

\subsection{Quantum ratio   \label{Qratio} }   

We have seen above  (Sec.~\ref{basic}, Sec.~\ref{temperature})   that several factors cooperate in determining that an isolated macroscopic body  behaves classically, i.e., with a well-defined position and momentum at each instance.   They are the nonzero temperature (self-decoherence - mixture),  the large mass or a large number of atoms and molecules composing it   (suppressing the quantum diffusion), the macroscopic scales of  precision versus the universal Heisenberg's uncertainty limits, and so on. A macroscopic body composed of $N \sim 10^3, 10^{25},  10^{50}$, etc.  constituent atoms and molecules may further have vastly different internal structures, densities, temperature-dependent specific heats, mass densities and charge distributions, etc.  

 One wonders whether there exists an approximate but simple, universal criterion to tell whether an isolated body (small or large)  is classical or quantum mechanical.   We propose here a simple criterion involving what may be called quantum ratio $Q$,   (\ref{quantum}) and (\ref{classical}) below. 

Let us first introduce the concept of the quantum fluctuation range, $R_q$.  Basically it simply represents the linear extension (spatial support) of the pure-state  wave function describing the system.  In the case of a single particle, the normalization condition  $|| \psi ||  =1$   would tell that  the support is in general limited.
But if the particle is approximately a  momentum eigenstate, the extension $R_q$ can be arbitrarily large. 
 As for many-body systems, $R_q$ is the range over which the  pure-state wave function description is valid.  Note that there is no a priori upper limit in quantum mechanics \footnote{It was  noted  in \cite{KK} that this is due to the fact that QM does not contain any fundamental constant with the dimension of a length.  } over which the quantum correlations - entanglement - between two or more particles (spins, polarizations, etc.)  persist,  which lead to the fascinating phenomenon of quantum nonlocality.

 $R_q$  depends on the  environment-induced decoherence effects, or for an isolated particle, on its body temperature $T$  and on other details of the body, as  in (\ref{DecoMacro}),  (\ref{MacroT}).  $R_q$ depends also on time.   
  Decoherence severely reduces  $R_q$.
  
Another relevant concept is the size $L_0$ of the body.  $L_0$ is by definition the spatial support of the wave function describing the bound state. $L_0$ varies from $0$  for the elementary particles,   a few to several angstroms  for atoms and molecules, hundreds  or thousands of angstrom  for large molecules, 
a tenth to a few microns  for a virus or bacteria,  
 to    $O(10^{-3}\, cm)$    to  $ \sim O(10^{20}\, cm)$  of any macroscopic  bodies to  astrophysical bodies.

 Now consider an isolated body (or a particle) at finite body temperature.  If its CM has the quantum range $R_q$ and its size is $L_0$,  our criterion is that  
 its CM  will behave, at distances larger than $L_0$,  quantum-mechanically or classically,  according to whether   
 \be      Q   \equiv   R_q/ L_0  \gg 1\;, \qquad  {\rm (quantum)},
 \label{quantum}   \ee
 or 
  \be      Q =  R_q /L_0  \lesssim  1 \;,\qquad  {\rm (classical)},     \label{classical} 
 \ee
 respectively. 
 
As already noted,  the elementary particles have  $L_0=0$  \footnote{The elementary particles  known today are  the leptons (electron, muon, $\tau$ lepton),   the neutrinos, the quarks,  $W$, $Z$ bosons and the photon.
  The fact that they are very accurately described by the local quantum field theory - the standard   
   $    SU(3)_{QCD}\times  \{  SU(2)_L\times U(1)\}_{GWS}  $   gauge theory
 (GWS stands for Glashow-Weinberg-Salam, and QCD for  Quantum Chromodynamics)
 of the fundamental interactions  \cite{Weinberg}-\cite{GellMann} up to the energy scales of  $O(10)\, TeV$,     means    $L_0   \lesssim  O(10^{-18})\, cm$.  
 In future these  elementary particles might  turn out to be bound states of some constituents unknown today.
}, that is, $Q=\infty$.  They are quantum mechanical. 
 Note that by definition elementary particles have no internal excitation modes, thus there is no meaning in  ``body temperature" for them. 
 
 Atoms and molecules are quantum-mechanical bound states, made of atomic nuclei and electrons for the former, and 
 of atoms and electrons, for the latter.  They have well-understood composite structures and varieties of internal excitation modes. They have well-defined size, $L_0$, varying from $0.5 $ \AA  \,\,    to hundreds of \AA.   At distances  much larger than $L_0$,  they  behave as pointlike particles.  
 The atomic nuclei made of protons and neutrons are more strongly bound (by the strong or nuclear forces), and have smaller size  $L_0$  of the order of Fermi $\sim 10^{-13}$ cm.

 As long as 
 self- or environment-induced decoherence is suppressed,  they behave perfectly as quantum-mechanical particles,  with characteristic phenomena such as diffusion, diffraction, interference, quantum entanglement and quantum nonlocality.   The experimental studies with  $C_{60}$ or $C_{70}$ molecules already cited several times  \cite{C60,C70} are  good examples illustrating these points.  
 
 Classical behavior  of (typically macroscopic) bodies  is  characterized by the fact that the quantum fluctuations of their CM   
do not extend beyond their size, (\ref{classical}), even if they are  {\it   all}, ultimately, quantum-mechanical bound states. 
   
 The ratio  (\ref{quantum}) for quantum mechanical particles, can be stated with a slightly different nuance and emphasis,  in the context of  the general quantum measurement problems.  According to  \cite{KK},   what characterizes quantum mechanics is the fact  that the basic building blocks 
 (i.e., degrees of freedom)  of our world are particles, elementary or composite.  The elementary interactions are  spacetime  local processes.
 This is why  any quantum measurement process, in its heart,  is a {\it  spacetime pointlike event}. And this, in turn, explains the apparent ``wave function collapse", the fact that the aftermath of each measurement corresponds to one particular  term of the orginal wave function describing the microscopic system being measured. The environment-induced, decohered, classical behavior of the measuring-recording device, does the rest.  
  
 These reflections can be pushed  even deeper.  Indeed,  (non-relativistic) standard  quantum mechanics 
 is based on relativistic quantum field theories  (QFT),  theories of elementary particles.  It is well known that  the dynamical causality as well as spin-statistic connection of   the standard quantum mechanics can be  established only within the  QFT framework  (see  \cite{KK,KKGP} for more comments and references).

To elaborate further this important concept of quantum ratio,  defining it  more rigorously and studying it quantitatively in concrete examples,  is left for 
a separate work.

\subsection{Quantum and classical time evolutions \label{Evolution} }

We have derived above Newton's equations from the  Schr\"odinger equation, the Ehrenfest theorem.  
Newton's equation is usually considered as a second-order differential equation in time, whereas  the Schr\"odinger equation is a first-order equation. 
How can the former appear as an approximation to the latter?   

Actually, even in the classical mechanics,  the evolution equations are first order differential equations in the canonical formalism where the variables $x$ and $p$ are treated as independent: their relation emerges from the equation of motion. Thus there are no differences of principle between the classical and quantum evolution equations, from this point of view:  both are perfectly deterministic evolution systems \footnote{Here  we set aside the interesting phenomena of caos, both in the clasical and quantum context.}.  

The first important difference lies in the way  the initial condition is imposed at time $t=0$.  In classical mechanics, it is 
just the simultaneous values of $x(0), p(0)$ or a collection of these for a many-body problem.  It is represented by a point in the so-called phase space. 
Their values at any later times are determined uniquely by Newton's equations. 

In quantum mechanics,  the initial condition is the wave function $|\psi(t)\ckt$ at $t=0$.    It contains, in a sense, an infinite amount of information
(the relative frequencies ${\cal P}_n$  for the measurement of  {\it any}   variable  are completely determined by it).    One of the most remarkable features of QM  is the fact that, nevertheless,    $|\psi(0)\ckt$ can be determined (or prepared) completely by performing a finite number of simultaneous measurements of compatible variables.  Given {\it any}  system,  there is always such a finite, maximal set of observables (though their choice is not unique). 
 Therefore,   the difference between classical and quantum mechanics,  in  the way to set up the initial condition,  is indeed  substantial,  but not essential, from the logical point of view.

The   fundamental difference between classical and quantum mechanics lies 
  in  the distinct nature of their predictions,  once the state at time $t$   has been found,   either by Newton's or  Schr\"odinger's  equation.  In classical mechanics, the predicted values $x(t), p(t)$ themselves are what the experimentalists measure, so there is no subtlety there.

In QM,  given the initial condition $|\psi(0)\ckt$,  the wave function $|\psi(t)\ckt$ is uniquely  determined 
from  the  Schr\"odinger equation:
 the time evolution is fully deterministic in QM also. 
 But what can be predicted is only  the {\it  relative frequencies}  ${\cal P}_n$ for finding various outcome $F= f_n$, for {\it  any} variable, $F$.  
 
 What essentially distinguishes the time evolution in quantum mechanics  from that in classical mechanics, is the fact that  the quantity that evolves deterministically in time (the wave function)  is not an observable, i.e.,  not itself a physical, measurable quantity, even though it encodes  the information
 about  the relative frequencies for different outcomes, for  {\it  any} possible measurement at time $t$.

In author's view,  this is the key to understand how the two apparently contradictory properties of quantum mechanics,  the quantum nonlocality (quantum entanglement among particles which are no longer causally connected) with all those fascinating consequences, on the one hand, and the rigorous locality and causality \footnote{The proof of this fact requires working in the framework of relativistic quantum field theory  \cite{Peskin}.},   on the other,      can coexist in quantum mechanics .  

In the so-called hidden-variable models, supposed to represent  a more complete description of Nature than QM,    
 the statistical aspect of quantum mechanics is replaced by a classical statistical  (unknown)  distribution of the hidden parameters  $\{\lambda\}$ \cite{Bell}.
  Once the value(s)  of  $\{\lambda\}$ at $t=0$ is (are) chosen,  all possible  imaginable future experimental outcomes are  determined uniquely:
    it represents  classical evolutions.
    
A  (hidden-variable) theory,  having a logical structure so sharply different from that of quantum mechanics,  cannot explain  {\it  all} of the predictions of the latter. 
It could, actually,  if one is willing to give up local causality.   This is seen clearly, e.g., in  the unphysical, noncausal  behavior of the trajectory $X(t)$ in Bohm's pilot-wave theory \cite{Bell},\cite{KKGP}   which is a particular example of  hidden-variable model, designed by construction  to reproduce all QM predictions. 

If we insist, instead, that a hidden-variable model must respect locality and causality, {\it then}  its predictions necessarily satisfy certain inequalities  such as those  
 in \cite{Bell}, \cite{CHSH}.     Bell and  Clauser et. al. have shown that {\it  some}  predictions  of QM  will necessarily lie outside those bounds,  a fact  verified in a dramatic experimental confirmation of QM by Aspect and others \cite{Aspect}.

\section{Summary and discussion \label{Summary}} 

In this work we have derived Newton's (force-) law from the Schr\"odinger equation \footnote{Newton's other law's,  that of inertia (the first law) and that of action-reaction
(the third law) are incorporated in QM from the beginning.}.

We first  identified the basic mechanism for the CM of a macroscopic body in the vacuum to possess a classical trajectory, i.e., a well-defined position and momentum at each instant of time. 
The first is the Heisenberg uncertainty relation.  A macroscopic body of $N\sim 10^{51}$ atoms satisfies, for its center of mass (CM),   the same Heisenberg
lower limit    for the product of the canonical position and momentum,  $\Delta X_i  \cdot   \Delta P_j  \ge  \delta_{ij}  \,  \hbar/2$ as for an atom.  
This means that for a macroscopic body, and for the measurements within the  appropriate range  of  macroscopic scale and precision,   Heisenberg's relations can be considered irrelevant.
The position and momentum can be measured simultaneously, to ``arbitrary" precision,   $\Delta X \approx 0$, $\Delta P\approx 0$.

The second is the absence of the diffusion for free wave packets.  Here the essential factor is the mass.  While for a free electron with mass $\sim 10^{-27} $ g,   the time needed for doubling its wave packet size  (starting from a $ 1   \, \mu m$)    is  $10^{-8}$ sec,    the corresponding time for a macroscopic 
body of $1$ g,  with the  initial CM wave packet of the same size $ 1   \, \mu m$,  exceeds   the age of the universe.   
This means that once its position and momentum are  experimentally measured with ``arbitrary"  precision,
       their uncertainties do not grow in time: a macroscopic body follows a 
classical (well-defined)  path:   a trajectory.   

 The third, and the crucial,  factor  is the nonzero temperature of the body, which makes it a radiating, metastable state.  It is a mixed state.   {\it  Decoherence}  caused by the entanglement with the ``lost"  (unobserved) photons  it emits and which carry away information,  means  that a split wave packet for a macroscopic body - 
 a coherent superposition of two macroscopically distinct state vectors   -    is a meaningless notion.  It cannot be prepared experimentally.

 Once  it is established that  the CM of an isolated macroscopic body at finite temperature has a unique classical trajectory,    
  the derivation of Newton's equation  for it is,  in principle, a straightforward application of Ehrenfest's theorem,  even though a special care is required (Sec.~\ref{subtle}).
    We have studied the case of weak gravitational forces  (Sec.~\ref{special1} and Sec.~\ref{general1}),  
a harmonic-oscillator potential (Sec.~\ref{special2}), 
and weak and almost space-independent external electromagnetic fields (Sec.~\ref{EM}).   
The derivation  of the canonical equations for many (macro-) body problems  has also been discussed  in Sec.~\ref{canonical}.

In Sec.~\ref{QCL}, we discussed critically the boundaries between the classical and quantum descriptions of the world, based on the 
analyses of Sec.~\ref{basic} and Sec.~\ref{Newton}.   
  First, we have discussed the boundary body temperatures, 
   drawing lessons from  several existing experimental studies.  They  show    that  the  rough estimate of the boundary temperature  is given by a simple formula, (\ref{temp}). It captures essentially the nature (and  the energy)  of the lowest  excitation  modes, characteristic of the system considered, such as  an atomic or molecular excitation, phonon excitations in a solid, or simply a macroscopic oscillation of the body.  The formula   (\ref{temp})  was found basically applicable to  mesoscopic-macroscopic bodies with number of atoms (or  of  molecules)  in a vast range spanning  from $N = 70$ to $N\sim  10^{27}$. 
 
 In Sec.~\ref{Qratio}  we have introduced the concept of quantum ratio.  We proposed an approximate, but universal criterion 
 (\ref{quantum}),(\ref{classical}),  
 to tell whether a  particle (a body)  behaves  quantum mechanically or classically.   
In Sec.~\ref{Evolution}  the distinct natures of the time evolutions in quantum and classical mechanics are critically revisited and discussed.

 Our way of  understanding  how  the classical mechanics emerges  from QM  for an isolated macroscopic body,  points to a  new perspective on the relation between quantum and classical physics.
  Namely, for a macroscopic body at finite temperatures  (hence in a mixed state),  classical mechanics for its center-of-mass motion emerges as   {\it   the first  approximation to 
  the quantum mechanical equations.}  \footnote{The content of Sec.~\ref{basic} and Sec.~\ref{Newton}  defines the approximations made to derive the classical equations.  The nature of the approximations involved there suggests extremely small corrections in general; 
  their magnitude will however depend on the kind of materials, mass,  and temperatures, on the presence of the environment and so on.  The study of quantum corrections to Newton's equations is a subject of future studies.   }

 This seems  to be a  novel point of view,  as  compared to  the traditional thinking about this problem,  for instance,    that  classical physics emerges in the so-called semi-classical approximation of QM  (i.e., in the large-action limit),    or that   the so-called coherent states  are the closest analogue of classical  objects  in QM.   Other textbook examples include   the formal similarity between the commutation relations in QM and the Poisson brackets in classical mechanics, and the fact  that  in semiclassical limit the Schr\"odinger equation for the (exponent of)  the wave function reduces to the 
  Hamilton-Jacobi equation. Still further examples are   the  Bohr-Sommerfeld quantization conditions and  Bohr's correspondence principle. 
    Though these well-known ideas  certainly represent  necessary conditions for the  QM to be compatible with classical mechanics, 
   and in that sense, are manifestations of  deep consistencies within QM,    
    they mostly concern pure quantum states.  None of these,  therefore,  represents the true  explanation of the emergence of  classical physics from QM,   
    which  hinges upon the  {\it  mixed-state, large mass}  nature of a macroscopic body {\it  at finite temperatures},      as we have seen 
     in the present work.

Also,   the quantum-classical  ``transition" discussed in this work concerns  a change {\it   in the way we describe the world} ; it does not deal with real discontinuities in physics.  The finite temperatures of a body are just signs of the body being in an excited, metastable quantum state. Any such system radiates spontaneously, and the entanglement with the photons it emits (and which carry away information)  render the system a mixture. 
Decoherence which results  is one of the main causes of the body to behave classically, as we saw.
 These considerations illustrate  clearly that the question of the emergence of the classical physics from QM examined in the present work
 is conceptually distinct from, and should not be confused with,    the physics of critical phenomena and various phase transitions, classical or quantum, which deal with an abrupt change in the collective rearrangement of microscopic degrees of freedom at macroscopic scales.

 Clearly the present work is in spirit  consistent with the idea of environment-induced decoherence  \cite{Joos1}-\cite{Arndt1}, as the key for emergence of classical world from quantum  mechanics.   It complements and strengthens this idea,  however, by clarifying  better the conditions for CM of a macroscopic body to have a classical trajectory, and 
 by explicitly deriving Newton's equations for some particular forces. We believe that the approximation and drastic simplification adopted here, 
  of considering an isolated macroscopic body in the vacuum, is adequate to extract simple laws of Nature,   Newton's classical mechanics, from QM.  The processes considered in proving environment-induced  decoherence constitute, instead, complicated backgrounds and corrections  from this point of view,  to be neglected in the first approximation.  Making such a simplification and idealization  (ignoring the air resistance, hence imagining the system in the vacuum) was implicit in Galileo's Gedankenexperiment argument \footnote{This theoretical argument characteristic of Galileo, is
  perhaps less known than the famous (actual or imaginary) experiment done from the top of the Leaning Tower, but is equally striking, to author's view. 
  },   that a light and a heavy stones must fall with the same rate, as otherwise we would be led to  an inconsistency if the two stones were bound and let fall together \cite{Galileo}.

\section*{Acknowledgments}

The inspiration for the present work came from an experiment performed in Pisa some $\sim15$ years ago,  by C. Bemporad, C. Bradaschia and F. Fidecaro,   
in a  celebrative event  for the public including school children.
 The setup consisted of two  $\sim3m$ tall  glass tubes with diameter $\sim 30$ cm, with their interior  kept in a good vacuum.   A few-centimeter steel  ball in the one and a light plastic sponge in the other, were let to fall simultaneously,   demonstrating that they indeed fall together (!),   evoking Galileo's  experiment  around the year  $\sim1590$.  
The device  
 is conserved in the entrance hall of the VIRGO/EGO  gravitational-wave observatory at Cascina, near Pisa. The author is deeply indebted  to F. Fidecaro for rearranging the reactivation of the  apparatus  and for demonstrating  this glass-tube ``Galileian" experiment  anew  for him. 
He  also  thanks  
F. Fidecaro and G. Cella for discussions on the physics of the gravitational wave detections  at LIGO and VIRGO.  
He thanks   P. Menotti and S. Bolognesi  for comments on the manuscript and for discussions.  A special thanks of the author is reserved for Hans Thomas Elze  for  many interesting 
discussions,  suggestions, and for useful information on the literature. 
   This work is supported by  the INFN special research initiative grant, ``GAST"  (Gauge and String Theories).

{}

\appendix

\section{Internal coordinates \label{coordinates} }

We introduce the (symmetric)  internal coordinates ${\hat  {\bf r}}_i$  as  
\be    {\bf r}_1 = {\bf R} +  c \,  {\hat {\bf r}}_1\;; \quad  {\bf r}_2 = {\bf R} + c \, {\hat  {\bf r}}_2\;; \quad \ldots,   \quad    {\bf r}_N = {\bf R} + c \, {\hat {\bf r}}_N\;;    \label{symmetric} 
\ee
($c$ is a constant, see below), where    ${\bf R}$ is the CM position and 
\be     \sum_{i=1}^N  m_i   {\hat {\bf r}}_i= 0\;. \label{relative2Ap} 
\ee
Any $N-1$ of   $ {\hat {\bf r}}_i $'s can be chosen to be linearly  independent.  
For definiteness,  we take $ {\hat {\bf r}}_1,  {\hat {\bf r}}_2, \ldots   {\hat {\bf r}}_{N-1}$ as the independent variables. Then the  choice of the constant $c$
\be    c= \left(\frac {m_N}{M}\right)^{\tfrac{1}{N-1}} \;,     \qquad  M \equiv   \sum_{i=1}^N m_i\;,  \label{Jacobian}
\ee
sets  the Jacobian of the coordinate  tranformation
\be     \{ {\bf r}_1,   {\bf r}_2,    \ldots     {\bf r}_N  \}  \longleftrightarrow      \{ {\bf R},   {\hat {\bf  r}}_1, \ldots      {\hat {\bf  r}}_{N-1} \} 
\ee
to unity, as can be easily verified.

Inverting  (\ref{symmetric}),  one has
\be         {\bf R} =  \frac{1}{M}    \sum_{i=1}^N    m_i   {\bf r}_i\;;\quad    {\hat {\bf r}}_1 =   \frac{1}{c}  ({\bf r}_1   - {\bf R})  \;; \quad \ldots ,    \quad   {\hat {\bf r}}_{N-1} = \frac{1}{c}  ( {\bf r}_{N-1}  - {\bf R}) \;.\label{rather} 
\ee
This leads to, by the standard chain rule,  
\bea    \nabla_1 &=&       \frac{m_1}{M}  \nabla_{\bf R} +  \frac{1}{c}  (1  -    \frac{m_1}{M} )   \nabla_{\hat 1}   -  \frac{m_1}{ c  M}    \nabla_{\hat 2}    \ldots   -    \frac{m_{1} }{c  M}    \nabla_{\widehat {N-1}} \;;
  \nonumber   \\
   \nabla_2  &=&        \frac{m_2}{M}  \nabla_{\bf R} -     \frac{m_2}{c M}    \nabla_{\hat 1}  +    \frac{1}{c}   (1  -  \frac{m_2}{M} )   \nabla_{\hat 2}   -  \ldots   -  \frac{m_{2} }{ c M}    \nabla_{\widehat {N-1}} \;;   \nonumber    \\
&&
    \ldots \ldots  \nonumber 
\\
    \nabla_i   &=&       \frac{m_i}{M}  \nabla_{\bf R} -    \frac{m_i}{c M}    \nabla_{\hat 1}  -  \ldots   +   \frac{1}{c}    (1  -   \frac{m_i}{M} )   \nabla_{\hat i}   - \ldots  -     \frac{m_{i} }{c  M}    \nabla_{\widehat {N-1}} \;;\nonumber
\\
  &&      \ldots \ldots  \nonumber 
\\
    \nabla_{N-1}   &=&    \frac{m_{N-1} }{   M}  \nabla_{\bf R}   -   \frac{m_{N-1}}{c  M}    \nabla_{\hat 1}  -  \ldots -    \frac{m_{N-1} }{c  M}    \nabla_{\widehat  {N-2}}  + \frac{1}{c}   (1  -  \frac{m_{N-1}}{M} ) \nabla_{\widehat {N-1}}     \;;\nonumber   \\
    \nabla_{N} &=&  \frac{m_N}{  M}  \nabla_{\bf R} -   \frac{m_{N}}  {c  M}    \nabla_{\hat 1}  -     \ldots  -    \frac{m_{N}}{c  M}   \nabla_{\widehat {N-1}}   \;.\label{nablas}
\eea
One sees that 
\be       \sum_{i=1}^N    \nabla_i=    \nabla_{\bf R}\;,    \qquad    {\bf P}=       \sum_{i=1}^N   {\bf p}_i  \label{momenta}
\ee
upon inspection of (\ref{nablas}).  
It is also  straightforward to check that  
\be   \sum_{i=1}^N    \frac{ \nabla_i^2 }{2 m_i}   =      \frac{ \nabla_{\bf R}^2 }{2 M}    +   \sum_{i=1}^{N-1}    \frac{ \nabla_{\hat  i}^2 }{2 \mu_i}  -  \frac{1}{c^2  M }  \sum_{i<j} 
 \nabla_{\hat  i} \cdot   \nabla_{\hat  j}  \label{nsimple} 
\ee
where  $\mu_i$  is the reduced mass, defined by 
\footnote{In two-body  ($N=2$) systems,  our internal coordinate coincides with the standard relative coordinate. Indeed,    $ {\hat  {\bf r}}_1 = {\mathbf r}_1 - {\mathbf r}_2$,   with   
$1/\mu_1=    M/ m_1 m_2$.
}     
\be    \frac{1}{\mu_i}   =  \left(\frac{M}{m_N}\right)^{{2}/{(N-1)}}    \left(  \frac{1}{m_i} - \frac{1}{M} \right) \;,\qquad      i=1,2,\ldots, N-1\;.   \label{reducedmass}
\ee
Actually,   for macroscopic bodies with $N \gg  1$,   $M \gg  m_i$, $M \sim O(N) m_N$,    which are the  focus of our interest in this work,     the constant $c$ and the reduced masses $\mu_i$'s, can be safely approximated by  
\be   c = 1\;,     \qquad         \forall i\;, \quad   \mu_i   =     m_i\;.   \label{good}
\ee

Even though the ``symmetric internal coordinates"   ${\hat {\bf r}}_i$    (\ref{symmetric})   allow the separation of the CM kinetic term from the  kinetic terms of the internal degrees freedom,  
the latter has not a simple form, (\ref{nsimple}).   This problem is solved by using the Jacobi coordinates  (also indicated below, as  ${\hat {\bf r}}_i $)  \footnote{They are just the relative coordinate between the CM  of the first $j$  particles and  ${\bf  r}_{j+1}$, for $j=1,2,\ldots, N-1$.   For $j=N$ it coincides with the total CM position.}  
\bea   {\hat  {\bf r}}_j   &= &     \frac{1}{m_{0 j}} \sum_{k=1}^j    m_k \,{\bf r}_k  -  {\bf r}_{j+1}  \;, \qquad  j=1,2,\ldots, N-1\;,     \nonumber \\
      {\hat {\bf r}}_N  &= &    \frac{1}{m_{0 N}} \sum_{k=1}^N    m_k \, {\bf r}_k  =  {\bf R} \;,   \qquad 
      m_{0 j}  \equiv     \sum_{k=1}^j m_k\;. 
\eea   
It is a simple matter to work out the linear transformation between  the independent variables in the two coordinate systems (the CM position ${\bf R}$ is common):
\be    \{   {\hat {\bf r}}_1,  {\hat {\bf r}}_2,\ldots  {\hat {\bf r}}_{N-1}\}^{\rm symmetric}   \longleftrightarrow    \{   {\hat {\bf r}}_1,  {\hat {\bf r}}_2,\ldots  {\hat {\bf r}}_{N-1}\}^{\rm Jacobi}\;. 
\ee
  In the Jacobi coordinates  the kinetic terms separate  as 
  \be    \sum_{i=1}^N  \frac {{\bf p}_i^2}{2m_i}  =            \frac{{\bf P}^2}{2M} +   \sum_{i=1}^{N-1}   \frac{ {\hat  p}_i^2} {2 \mu_i}\;,  
\qquad 
    {\bf P}  \equiv  - i \hbar \nabla_{\bf R}  \;, \qquad     {\hat  p}_i \equiv   -i \hbar    \nabla_{\hat  i}\;.      \label{simpleKin}
\ee
($1/\mu_i \equiv  1/m_{0j}+1/m_{j+1}$  here).

The relation (\ref{momenta}) holds true in  both systems. 
In the main text we may use either the Jacobi or symmetric coordinates  \footnote{The simple form of the kinetic term for the internal degrees of freedom such as 
  (\ref{Hamiltonians}),   (\ref{HamiltoniansBis}),  (\ref{HamiltoniansAgain}),   is thus valid in the Jacobi coordinates. Most of the discussions of this work  however depend only on  the separation of the CM Hamiltonian from the internal Hamiltonian describing the bound state.}.  For some discussions (such as in (\ref{quadr}))   the symmetric  internal coordinates appear  to be more convenient.   A side remark  is that  in atomic systems, 
where the center of mass approximately coincides with the position of the nucleus,  ${\bf r}_N\simeq  {\bf R}$,  the symmetric  coordinates (\ref{symmetric})
correspond to the independent position operators of the electrons in the rest frame of the atom.  Also the last term  on the r.h.s. of  (\ref{nsimple}) is negligible for atoms,  as $(N-1) \,m_e \ll  M \sim m_{nucleus}$, and the simple form of the kinetic terms (\ref{simpleKin})  is valid with $\mu_i=m_i$.

If the  potential has a simple separable form, 
\be     V ({\bf r}_1,  {\bf r}_2, \ldots  {\bf r}_N) =      V({\mathbf R}) +    V_{int}({\hat  {\bf r}}_1,  {\hat  {\bf r}}_2, \ldots   {\hat  {\bf r}}_{N-1}) \;,
\ee
then   the total Hamiltonian  simplifies as   
\be  H= H_{CM} +   H_{int}\;,   
\label{TotHamAgain}  \ee 
\be       H_{CM}=  \frac{{\bf P}^2}{2M} +  V({\mathbf R})\;;  
  \qquad     H_{int}= \sum_{i=1}^{N-1}    \frac{{\hat p}_i^2}{2\mu_i} +   V_{int}({\hat {\bf r}}_1,  {\hat {\bf r}}_2, \ldots   {\hat {\bf r}}_{N-1}) \;.
\label{HamiltoniansAgain}  \ee
Such a  Hamiltonian allows us to write the wave function in a factorized form, 
\be   \Psi({\bf r}_1,  {\bf r}_2, \ldots  {\bf r}_N) =   \Psi_{CM}({\bf R}) \psi({\hat  {\bf r}}_1,  {\hat  {\bf r}}_2, \ldots   {\hat  {\bf r}}_{N-1}) \;.   \label{factorizationBis}
\ee

\section{Perturbation theory \label{PertEh} }

In the discussions of Sec.~\ref{general1} a question arose how  the perturbation theory can be applied to find out the 
way    Ehrenfest's theorem gets affected by a perturbation.  In this Appendix,  we use  the standard  (time-independent) perturbation theory  for a 1D particle,    to  work out   the
question.  For simplicity, we consider  a system  with a nondegenerate, discrete spectrum, and with the perturbative potential depending only on $x$.

The Hamiltonian has the form, 
\be   H = H_0  +  \lambda V^{\prime}(x)  \;,   \label{Pert}  
\ee
where $H_0$ is the unperturbed Hamiltonian,  
\be   H_0 =   \frac{p^2}{2m}  +  V_0\;;    
\ee
$ V^{\prime}$  \footnote{``prime" is there to indicate a perturbation, not a derivative.}     represents the perturbation. An arbitrary parameter $\lambda$  (which can be set to $1$  at the end) has been introduced to  keep 
track of   the power of $V^{\prime}$  in various expressions.

The problem is to find
the power-series solution for   $E(\lambda)$ and 
$\ket{\psi(\lambda)}$ 
\begin{align}
\ket\psi &= \ket{\psi^{(0)}} +\lambda \, \ket{\psi^{(1)}}+\lambda^2 \,\ket{\psi^{(2)}}+\ldots + \lambda^{L} \ket{\psi^{(L)} }+\ldots;  \label{lez1.4}\\
E &= E^{(0)} + \lambda \,\varepsilon_1 + \lambda^2 \, \varepsilon_2+\ldots.     \label{expansion} 
\end{align}
such that 
\be 
H(\lambda)\,  \ket{\psi(\lambda)} = E(\lambda) \,  \ket{\psi(\lambda)}\;,\label{eq1}\ee
order by order in $\lambda$,    and 
with the boundary  condition
\be
\lim_{\lambda\to 0} E(\lambda) = E^{(0)} \qquad   \lim_{\lambda\to 0}\ket{\psi(\lambda)}
= \ket{\psi^{(0)}}\;.       \label{initial}\ee
By denoting the unperturbed solutions  of  
\be     H_0  \ket{\psi^{(0)}} =  E^{(0)} \ket{\psi^{(0)}} \;, 
\ee
as
\be     H_0  \ket{k} =  E_k^{(0)} \ket{k} \;, \quad k=0,1,2,\ldots 
\ee
the problem of finding the corrections to the  $n$-th energy eigenvalue and   associated eigenstate  can be formulated by taking
\be       \ket{\psi^{(0)}}= \ket {n}\;, \qquad     E^{(0)}=    E_n^{(0)}   \label{boundary}
\ee
in (\ref{expansion}) and (\ref{initial}) and solving for    $\ket{\psi^{(L)}}$ and $\epsilon^{(L)} $.  $\lambda$ can be set to $1$ after the calculation.  Because of the boundary condition, 
(\ref{boundary}), all the expressions for the corrections to the energy eigenvalue and wave function   below  should carry the index $``(n)"$; it is  however omitted as usual, for not cluttering the formulae. 

The results can be found in any QM textbook:  to first  order  they are given by :
\be \varepsilon_1=  \bra{\psi^{(0)}} V^{\prime} \ket{\psi^{(0)}}  =  \brc {n}| V^{\prime} | {n} \ckt = V^{\prime}_{nn} \; ,  \label{lez1.7a}\ee
\be   \ket{\psi^{(1)}} = \sum\nolimits'_{k} \ket{k} \frac{1}{E_{n}^{(0)}-E_{k}^{(0)}}\bra{k} V^{\prime} \ket{\psi^{(0)}}
= \sum\nolimits'_{k} \ket{k} \frac{V^{\prime}_{kn}}{E_{n}^{(0)}-E_{k}^{(0)}}
\;.
\label{lez1.7b}\ee
where the prime over the summation symbol means    $k \ne  n$.
The higher-order results can be found easily by an iterative procedure,  see for instance \cite{KKGP}. The results are
\be \varepsilon_L = \bra{\psi^{(0)}} V^{\prime} \ket{\psi^{(L-1)}}\;,\qquad L=1,2,3, \ldots \label{lez1.10}\ee
which determines the  correction to the energy to order $\lambda^L$, and   
\be \scalar{k}{\psi^{(L)}}  = \frac{1}{E_{n}^{(0)}- E_k^{(0)}}\bra k  V^{\prime} \ket{\psi^{(L-1)}} - \frac{1}{E_{n}^{(0)}- E_k^{(0)}}
\sum_{K=1}^{L-1} {\varepsilon_K} \scalar{k}{\psi^{(L-K)}} \;.\label{pt6}\ee
for the wave function.

Now let us consider the effects of perturbation  (\ref{Pert})  in Ehrenfest's theorem: 
\be   \frac{d}{dt}  \brc \psi   |  x |\psi \ckt  =   \frac{1} {i\hbar}   \brc \psi   |  [x, H]  |\psi \ckt =   \brc \psi  | \frac{p}{m}  |\psi \ckt  \;; \label{Ehr111}
\ee
\be   \frac{d}{dt}  \brc \psi   |  {p}  |\psi \ckt  =   \frac{1} {i\hbar}   \brc \psi   |  [{p}, H]  |\psi \ckt =   - \brc \psi  |   \frac{dV_0}{dx} +  \lambda   \frac{d
V^{\prime}}{dx}  |\psi \ckt  \;.  \label{Ehr222}
\ee
We note that for perturbations depending only on $x$,  (\ref{Pert}),  the  classical  relation 
\be     \frac{d  x_0(t)  }{dt}     =  \frac{p_0(t) }{m} \;, \qquad     x_0(t)\equiv   \brc \psi   |  x |\psi \ckt  \;;\quad  
  p_0(t)\equiv   \brc \psi   |  p   |\psi \ckt \;,   
\ee
holds to all orders of perturbation.  The second Ehrenfest's relation,   (\ref{Ehr222}),  also  contains on its left hand side 
$dp_0(t)/dt$  to all orders.    The right hand side of  (\ref{Ehr222})   thus    may be considered as an expression for  the force, $F$.    $F$ may be 
expanded as
\be    F=    F^{(0)} +  \lambda   F^{(1)}  + \ldots   +  \lambda^L   F^{(L)} + \ldots    \label{force1}
\ee
where 
\be        F^{(0)}  = -    \brc \psi^{(0)}   |   \frac{dV_0}{dx} |\psi^{(0)} \ckt  \;; \label{force00} 
\ee
\be   F^{(1)}  = -    \brc \psi^{(0)}   |   \frac{dV^{\prime}}{dx} |\psi^{(0)} \ckt -       \brc \psi^{(0)}   |   \frac{dV_0}{dx} |\psi^{1)} \ckt  -   \brc \psi^{(1)}   |   \frac{dV_0}{dx} |\psi^{(0)} \ckt\;, 
     \label{force01}
\ee
etc.    The  corrections to the force    to order  $L$  are  given by
\be       F^{(L)}  = -  \sum_{K=0}^{L}     \brc \psi^{(K)}   |   \frac{dV_0}{dx} |\psi^{(L-K)} \ckt  
 -  \sum_{K=0}^{L-1}     \brc \psi^{(K)}   |   \frac{dV^{\prime}}{dx} |\psi^{(L-K-1)} \ckt \;,\label{forceall}  
\ee
where the wave function corrections are given in  (\ref{pt6}).  

In view of the application in Sec.~ \ref{general1}, let us work out the second-order result fully.   
\bea      F^{(2)}  &=&   -   \brc \psi^{(0)}   |   \frac{dV_0}{dx} |\psi^{(2)} \ckt  -    \brc \psi^{(1)}   |   \frac{dV_0}{dx} |\psi^{(1)} \ckt  - \brc \psi^{(2)}   |   \frac{dV_0}{dx} |\psi^{(0)} \ckt  
\nonumber   \\
 &-&     \brc \psi^{(0)}   |   \frac{dV^{\prime}}{dx} |\psi^{(1)} \ckt   -    \brc \psi^{(1)}   |   \frac{dV^{\prime}}{dx} |\psi^{(0)} \ckt  \;,\label{force02}  
\eea
where
\be    |\psi^{(1)}\ckt =     \sum_{k\ne n} \ket{k} \frac{V^{\prime}_{kn}}{E_{n}^{(0)}-E_{k}^{(0)}}\;, 
    \qquad    \varepsilon_1= V^{\prime}_{nn} \;, 
        \label{wfcorr1}
\ee
  \be \scalar{k}{\psi^{(2)}}  = \frac{1}{E_{n}^{(0)}- E_k^{(0)}}\bra k  V^{\prime} \ket{\psi^{(1)}} - \frac{1}{E_{n}^{(0)}- E_k^{(0)}}
 {\varepsilon_1} \scalar{k}{\psi^{(1)}} \;.  \label{wfcorr2} \ee


\begin{thebibliography}{}


  \bibitem{Joos1} 
   E.~Joos and  H.~D.~Zeh, 
 ``The emergence of classical properties through interaction with the environment", 
 Z. Phys. B 59, 223-243 (1985).
 
   
  \bibitem{Zurek1}
  W.~H.~Zurek,
  ``Decoherence and the Transition from Quantum to Classical",
   Physics Today 44, 10, 36 (1991); 
    https://doi.org/10.1063/1.881293.
    
    
     \bibitem{Tegmark}  
 M.~Tegmark,
``Apparent wave function collapse caused by scattering,''
Found. Phys. Lett. \textbf{6}, 571 (1993)
doi:10.1007/BF00662807
[arXiv:gr-qc/9310032 [gr-qc]].



 \bibitem{Joos}  
  E.~Joos, H.~D.~Zeh, C.~Kiefer, D.~Giulini, J.~Kupsch, I.~O.~Stamatescu,
  ``Decoherence and the Appearance of a Classical World in Quantum Theory",
  Springer, 2002. 


 
    \bibitem{Zurek2}
  W.~H.~Zurek,
``Decoherence, einselection, and the quantum origins of the classical,''
Rev. Mod. Phys. \textbf{75}, 715-775 (2003)
doi:10.1103/RevModPhys.75.715
[arXiv:quant-ph/0105127 [quant-ph]].

 
\bibitem{Arndt1}
M.~Arndt and K.~Hornberger,
``Testing the limits of quantum mechanical superpositions", 
Nature Physics 10 271-277 (2014),
[arXiv:1410.0270v1 [quant-ph]].



\bibitem{WheelerZ}
 J.~A.~Wheeler,  W.~ H.~Zurek,   
  ``Quantum Theory and Measurement",
  Princeton NJ:  Princeton University Press (1983). 
   


\bibitem{Bell}  
J.~S.~Bell, 
``Speakable and unspeakable in Quantum Mechanics",
Cambridge University Press  (1987).


   
   \bibitem{Peres} 
  A.~Peres, 
  ``Quantum Theory: Concepts and Methods",   Dordrecht/Boston/London:  Kluwer (1995). 
  


\bibitem{KK}
K.~Konishi,
``Quantum fluctuations, particles and entanglement: a discussion towards the solution of the  quantum measurement problems,''
Int. Journ. Mod. Phys. A  37 (2022)  2250113
[arXiv:2111.14723 [quant-ph]].


\bibitem{KKTalk}
K.~Konishi,
``Quantum fluctuations, particles and entanglement: solving the  quantum measurement problems,''
to appear in proceedings of the conference DICE2022,
September 19-23, 2022 (Castiglioncello, Tuscany), Journ. Phys.: Conf. Series".   



\bibitem{Cornell}
 M. H. Anderson, J. R. Ensher, M. R. Matthews, C. E. Wieman, E. A. Cornell,
  Science 269, 198 (1995).

\bibitem{Ketterle}  
 K. B. Davis, M.-O. Mewes, M. R. Andrews, M. J. Van Druten, D. S. Durfee, D. M. Kurn, W. Ketterle, 
 Phys. Rev. Lett. 75, 3969 (1995).
 
\bibitem{Cohen} 
 F. Pereira Dos Santos, J. Léonard, J. Wang, C. Barrelet,
F. Perales, E. Rasel, C. Unnikrishnan, M. Leduc,
C. Cohen-Tannoudji, 
Phys. Rev. Lett. 86, 3459 (2001).

 


\bibitem{LL9} 
E. M. Lifshitz  and  L. P. Pitaevskii,    
    ``Statistical Physics, Part 2", 
  Landau and Lifshitz Course of Theoretical Physics, Vol. 9
(1980).

\bibitem{He3}
D.~D.~Osheroff, R.~C.~Richardson and D~.M.~ Lee,
``Evidence for a New Phase of Solid He3",  
 Physical Review Letters 28, 885 (1972). 
 
   \bibitem{C70}   
    L.~ Hackerm\"uller, K.~Hornberger,  B.~ Brezger, A.~Zeilinger, M.~Arndt, 
  ``Decoherence of matter waves by thermal emission of radiation",
  {\it Nature}, 427, (2004)  711. 
arXiv:quant-ph/0402146.


\bibitem{Clauser}
J.~F.~Clauser and S.~Li,
``Talbot-vonLau atom interferometry with cold slow potassium,''
Phys. Rev. A \textbf{49}, no.4, R2213-R2216 (1994)
doi:10.1103/physreva.49.r2213.

%
\bibitem{Hansen}
K.~Hansen and E.~E.~B.~Campbell,
``Thermal radiation from small particles", 
Phys. Rev. E   58 (1998),   5477. 




\bibitem{Birrell} 
N.~D.~Birrell, P.~C.~W.~Davies, 
``Quantum Fields in Curved Space", 
Cambridge University Press (1982). 



  
 \bibitem{SQL}
V.B. Braginsky, 
``Development of quantum measurement methods
(Methodological notes on part of Einstein's scientific legacy)",  
Physics-Uspekhi,   48  (2005),  595.


 

 \bibitem{GW150914}
B. P. Abbott et al.,
(LIGO Scientific Collaboration and Virgo Collaboration)
 ``Observation of Gravitational Waves from a Binary Black Hole Merger", 
Phys. Rev. Lett. 116, 061102 (2016). 


 
\bibitem{Braginsky}
V.~B.~Braginsky, M.~L.~Gorodetsky, F.~Y.~Khalili, A.~B.~Matsko, K.~S.~Thorne and S.~P.~Vyatchanin,
``The Noise in gravitational wave detectors and other classical force measurements is not influenced by test mass quantization,''
Phys. Rev. D \textbf{67}, 082001 (2003)
doi:10.1103/PhysRevD.67.082001
[arXiv:gr-qc/0109003 [gr-qc]].




\bibitem{Heisenberg} 
  W.~Heisenberg,
  ``The physical principles of the quantum theory", 
  Chicago University Press (1930), 
  Dover publications, INC. (1949).
  
  
   
  \bibitem{Tonomura} 
A.~ Tonomura, J.~Endo, T.~Matsuda, T.~Kawasaki and H.~Ezawa,
``Dimonstration of single-electron buildup of interference pattern", 
American Journal of Physics  57,  (1989) 117.

 



\bibitem{Peskin} 
M.~E.~Peskin and  D.~V.~Schr\"oder 
``An Introduction to Quantum Field Theory", 
 (New York, Addison-Wesley, 1995).




 
%
\bibitem{KKGP}
K.~Konishi and G.~Paffuti,
``Quantum Mechanics: A New Introduction", 
Oxford University Press  (2009).


\bibitem{CHSH}
J.~F. Clauser,  M.~A.~ Horne,  A.~Shimony, R.~A.~Holt,
 ``Proposed experiment to test local hidden-variable theories",
Phys. Rev. Lett., 23 (15): 880 
 (1969).  


\bibitem{Aspect}
 A.~ Aspect, P.~ Grangier, G. Roger,
 ``Experimental realization of Einstein-Podolsky-Rosen-Bohm Gedankenexperiment: A new violation of Bell's inequalities", 
 Phys. Rev. Lett.  49 (1981) 
91.  

%

    \bibitem{Leggett} 
  A.~J.~Leggett, 
  ``Macroscopic Quantum Systems and the Quantum Theory of Measurement",
  Suppl. of  the Progress of  Theoretical Physics,  69  (1980) 80. 
 
  
  \bibitem{C60}   
  M.~Arndt, O.~Nairz, J.~Vos-Andreae, C.~ Keller, G. van der Zouw, A.~Zeilinger, 
  ``Wave-particle duality of $C_{60}$  molecules",
  Nature  401 (1999) 680. 
  
  
  \bibitem{County}
  J.-M. Courty, A.~Heidmann, and M.~ Pinard,
  ``Quantum limits of cold damping with optomechanical coupling",
Eur. Phys. J. D 17, 399–408 (2001).
  
  \bibitem{Armour}
    A.~D.~ Armour, M.~P.~ Blencowe, K.~C.~ Schwab,
    ``Entanglement and decoherence of a Micromechanical Resonator via Coupling to a Cooper-Pair Box",
    Phys. Rev. Lett.  88 (2002), 148301. 
    
\bibitem{Knobel}    
R.~G.~Knobel, A.~N.~Cleland,
    ``Nanometre-scale displacement sensing using a single electron transistor",
    Nature,  424  (2003) 17.

    
      \bibitem{LaHaye}  
  M.~D.~LaHaye, O.~Buu, B.~Camarota, K.~C.~Schwab,
  ``Approaching the Quantum Limit of a Nanomechanical Resonator",  
  Science 304, 74–77 (2004).
  


  \bibitem{Cleland}
  A.~N.~Cleland, M.~R.~Geller, 
  ``Superconducting Qubit Storage and Entanglement with Nanomechanical Resonators",
  Phys. Rev. Lett.  93  (2004)  070501.
  
  \bibitem{Martin}
   I.~Martin, A.~Shnirman, L.~ Tian,  P.~ Zoller,
  ``Ground-state cooling of mechanical resonators",
Phys. Rev.  B 69, 125339   (2004).
  
  
  \bibitem{Kleckner}
  D.~ Kleckner,  D.~ Bouwmeester,
  ``Sub-kelvin optical cooling of a micromechanical resonator", 
Nature,  Vol 444|2 (2006).
  
  \bibitem{Regal}
  C.~A.~ Regal, J.~D.~Teufel,  K.~W.~Lehnert,
  ``Measuring nanomechanical motion with a microwave cavity interferometer", 
  Macmillan Publishers Limited  (2008),
  doi:10.1038/nphys974.
  
  \bibitem{Schliersser}
  A.~Schliesser, R.~Rivi\`ere, ,G.~Anetsberger, O.~Arcizetandt, J.~Kippenberg, 
  ``Resolved-sideband cooling of a micromechanical oscillator", 
  Macmillan Publishers Limited  (2008),
  doi:10.1038/nphys939.

  

\bibitem{LIGO}
B.~Abbott \textit{et al.} [LIGO Scientific],
``Observation of a kilogram-scale oscillator near its quantum ground state,''
New J. Phys. \textbf{11}, 073032 (2009)
doi:10.1088/1367-2630/11/7/073032.

  
  
    
 \bibitem{Aaron}
  A.~D.~O'Connell, et. al.,   
  ``Quantum ground state and single-photon control of a mechanical resonator",
  {\it Nature}, 464(7289)  (2010)  697.
 

  
  
\bibitem{Brand}
 C.~Brand, S.~Troyer, C.~Knobloch, O.~Cheshinovsky, M.~A.~Arndt,  
``Single, double and  triple-slit diffraction of molecular matter waves",
arXiv:2108.06565v2 [quant-ph],  2021.


\bibitem{Weinberg}    
S.~Weinberg,  
``A model of Leptons", 
{\it  Phys. Rev. Lett. } {\bf 19}, 1264  (1967).

\bibitem{Salam}

A~. Salam ``Weak and electromagnetic interactions,"  in “Elementary Particle Theory”, ed. N. Svartholm, Almqvist Forlag AB, 367
 (1968).
 
\bibitem{Glashow}
S~.L~. Glashow, J~. Iliopoulos and L~. Maiani (1970), 
``Weak Interactions with
Lepton-Hadron Symmetry,"  
 {\it Phys. Rev. }   D    {\bf 2}, 1285 (1970).  
 
 
 
 \bibitem{GellMann}
H.~Fritzsch, M.~Gell-Mann,  H.~Leutwyler,  
"Advantages of the color octet gluon picture",
{\it Physics Letters}  {\bf 47} B   (1973),  365.  



  \bibitem{Feynman}  
R. Feynman,   (1965),  
 ``The Character of Physical Law",
 The M.I.T. Press  (1965).  
   
  
  \bibitem{Landau6} 
  L.~D.~Landau and E.~M.~Lifshitz,  
  ``Fluid Mechanics", 
  Volume 6 of Course of Theoretical Physics,  Elsevier (1987).
  
     
  \bibitem{Tritton}
  D.~J.~Tritton,
  ``Physical Fluid Dynamics",  
  Oxford University Press (1988).
  

  
  
    \bibitem{Stewart}
  R.~H.~Stewart, 
  ``Introduction to Oceanology",    
Texas  A \& M University, (1997).   
  



\bibitem{Landau8}  
L.~D.~Landau and E.~M.~Lifshitz,  
``Electrodynamics of continuous media", 
Volume 8 of Course of Theoretical Physics,  Pergamon Press (1954).



\bibitem{Newton} 
I. Newton, 
 ``Philosophiae naturalis principia mathematica",
 Londini, iussu Societatis Regiae ac typis Josephi Streater, anno MDCLXXXVII   (1687). 
 
 
 \bibitem{Galileo}  
 G.~Galilei, 
 ``Dialogo sopra i due massimi sistemi del mondo"  (1632),
 Einaudi (1970). 



\end{thebibliography}
\end{document}